%% file: D1_note.tex
\documentclass[12pt]{article}
\usepackage{cite}
\usepackage[T1]{fontenc}
\usepackage[utf8]{inputenc}

\usepackage[english]{babel}

\usepackage{graphicx}
\usepackage[dvipsnames]{xcolor}
\usepackage{gensymb}
\usepackage{multirow}
\usepackage[thinlines]{easytable}
\usepackage{amsmath}
\usepackage{amsfonts}
\usepackage{amssymb}
\usepackage{amstext}
\usepackage{cancel}
 
\allowdisplaybreaks
\usepackage{slashed}
\usepackage{braket}
\usepackage{tablefootnote}
\usepackage[bookmarks, breaklinks, colorlinks,urlcolor=black, citecolor=red, 
linkcolor=blue]{hyperref}

\numberwithin{equation}{section}

\input{macros.tex}

\begin{document}

\begin{flushright}
SI-HEP-2022-04\\
P3H-22-027 \\[0.2cm]
\end{flushright}
	
\renewcommand*{\thefootnote}{\fnsymbol{footnote}}

\vskip 2cm	
	
	\begin{center}
		
		{\Large\bf \boldmath $B\to D_1(2420)$ and $B\to D_1'(2430)$ form factors \\[2mm]from QCD light-cone sum rules} \\[6mm]
	{
	Nico
	Gubernari\,\footnote{Email: nicogubernari@gmail.com}, 
	Alexander Khodjamirian\,\footnote{Email: khodjamirian@physik.uni-siegen.de}, 
    Rusa Mandal\,\footnote{Email: Rusa.Mandal@uni-siegen.de},
		and  Thomas Mannel\,\footnote{Email: mannel@physik.uni-siegen.de }
}
\\[6pt]
	
	{\small\it Center for Particle Physics Siegen (CPPS), Theoretische Physik 1,
           \\ Universit\"at Siegen, 57068 Siegen, Germany}

	\end{center}
	
	%

\begin{abstract}
\noindent
 We perform the first calculation of  form factors in the semileptonic decays
 $B\!\to\! D_1(2420)\ell\nu_\ell$ and $B \to D_1^\prime (2430)\ell \nu_\ell$  using QCD  light-cone sum rules (LCSRs) with $B$-meson distribution amplitudes. In this calculation 
 the $c$-quark mass is finite.
 Analytical expressions for two-particle contributions up to twist four are obtained. To 
 disentangle the $D_1$ and $D_1^\prime$ contributions in the LCSRs, we suggest a novel approach that introduces a combination of two  interpolating currents for these charmed mesons.
 To fix all the parameters in the LCSRs,
 we use the two-point QCD sum rules for the decay constants of $D_1$ and $D_1^\prime$ mesons 
 augmented
 by a single experimental input, that is  the $B \to D_1(2420)\ell\nu_\ell$ decay width. We 
 provide numerical results for all $B\to D_1$ and $B\to D_1^\prime$ form factors.
 As a byproduct, we also obtain the $D_1$- and $D_1'$-meson decay constants
 and predict the lepton-flavour universality ratios $R(D_1)$ and $R(D_1')$.

\end{abstract}

\renewcommand*{\thefootnote}{\arabic{footnote}}
\setcounter{footnote}{0}

\newpage

{\hypersetup{linkcolor=black}\tableofcontents}

\section{Introduction}

The $B\to D \ell \nu_\ell$ and $B\to D^*\ell\nu_\ell$ 
weak semileptonic decays  have been broadly studied in the last decades (for a recent review see, e.g., 
Ref.~\cite {Gambino:2020jvv}). 
The predictions of observables in these decays critically depend on our knowledge of the corresponding hadronic form factors.
The  $B\to D^{(*)}$ form factors  
are calculated in lattice QCD with a continuously increasing accuracy \cite{Aoki:2021kgd}. 
These  form factors can also be obtained from QCD  sum rules, where the most advanced
computations are in Refs.~\cite{Faller:2008tr,Wang:2017jow,Gubernari:2018wyi}.
In addition, in the heavy-mass limit --- that is for $m_c,m_b \to \infty$ ---
the $B\to D^{(*)}$ form factors are constrained and normalized at the zero recoil point 
corresponding to the maximal  momentum transfer to the leptons.

Less explored from both the theoretical and 
the experimental sides are the $B$-meson semileptonic transitions to the 
lowest lying excited 
charmed mesons  $D_0^*$, $D_1^{(\prime)}$, $D_2^*$ with the spin-parity $J^P=0^+,1^+,2^+$,
respectively. 
Nonetheless, an accurate knowledge of the form factors 
of these transitions is important for several reasons. 
Firstly, there is a long standing problem of filling
the gap between the inclusive $B\to X_c\ell \nu_\ell$ width 
and the sum of the exclusive semileptonic widths dominated by the $B \to D \ell\nu$ and $B \to D^* \ell\nu$ modes~\cite{pdg}. 
Secondly, the constraints on the $B\to D^{(*)}_{0,1,2}$ form factors from 
 the heavy-mass limit are weaker, since we do not have a
 normalization condition for these cases.  
 Finally, the  observed tension
 in the lepton-flavour universality (LFU) ratios $R(D)$ and $R(D^*)$ demands LFU tests  in all channels of 
the exclusive $b\to c$ transitions, including 
$B\to D^{(*)}_{0,1,2} \ell \nu_\ell$.
It is therefore extremely important to calculate the form factors 
of $B$ transitions to each of these excited charmed mesons.
\\

The masses and widths of the lowest four charm resonances with $J^P=0^+,1^+,2^+$ are collected in Table~\ref{tab:spectr}.
In the heavy mass limit for the charm quark, e.g. 
within Heavy Quark Effective Theory (HQET), one  indeed expects four states with one unit of orbital angular momentum. Since the heavy quark spin decouples,
these states fall into two (mass degenerate) spin-symmetry doublets, 
differing by the total angular momentum $j$ of light degrees of freedom.
One doublet with $j=1/2$ consists of $0^+$ and $1^+$ states and a second doublet with $j=3/2$  consists of $1^+$ and $2^+$ states.

In this work we focus on the two $1^+$ mesons $D_1$ and $D_1'$. 
Again, one invokes HQET where  the strong transition
$|j=1/2\rangle\to |j=1/2\rangle+\pi$ proceeds via $S$-wave whereas in 
the transition $|j=3/2\rangle \to |j=1/2\rangle +\pi$ only the $D$-wave is possible, causing a kinematical suppression. Hence,
the $j=1/2$ state has a 
significantly larger total width than the $j=3/2$ state 
\cite{Isgur:1991wq,Lu:1991px}. 
It is then natural to expect that the observed narrow  
$D_1$ (broad $D_1'$) resonance decaying to $D^* \pi$ is predominantly  
the $j=3/2$ ($j=1/2$) state. In reality,
a mixing between the two HQET states  inevitably takes place 
if one goes beyond the $m_c\to \infty$ limit.
 Note that the mixing pattern of $D_1$ and $D_1'$ 
in terms of HQET states is purely non-perturbative. It has been discussed in a model dependent framework
(see, e.g., Ref.~\cite{Klein:2015doa} for further details).
It is then desirable to calculate the $B\to D_1$ and
$B\to D_1'$
form factors in a finite $c$-quark mass framework.
\\

\begin{table}[t]
	\begin{center}
		\begin{tabular}{|c|c| c| c |c|}
			\hline \noalign{\vskip 2pt}
			Meson& $j$ & $J^P$ & Mass [\mev] & Width [\mev] \\ [.7ex]
			\hline\hline \noalign{\vskip2pt}
			$D_0^*(2300)$ & $\frac{1}{2}$ & $0^+$ & $2343\pm 10$ & $229\pm 16$ \\[1ex]
			$D_1(2430)\equiv D_1'$ & $\frac{1}{2}$ & $1^+$ & $2412\pm 9$ & $314\pm29$ \\[1ex] \hline\noalign{\vskip2pt}
			$D_1(2420)\equiv D_1$ & $\frac{3}{2}$ & $1^+$ & $2422.1\pm 0.6$ & $31.3\pm 1.9$\\[1ex]
			$D_2^*(2460)$ & $\frac{3}{2}$ & $2^+$ & $2461.1\pm 0.8$  & $47.3\pm 0.8$ \\[1ex]
			\hline
		\end{tabular}
		\caption{\emph{The lowest excited charmed mesons. For definiteness, we quote the masses and total widths of electrically neutral states from \cite{pdg}. 
		}
		\label{tab:spectr}
		}
		\end{center}
\end{table}

QCD sum rules \cite{Shifman:1978bx}
have been already used to evaluate the $D^{(\prime)}_1$ decay constant 
and the $B\to D^{(\prime)}_1$ form factors from two-point 
and three-point correlators, respectively.
This was done mostly in HQET (see, e.g., Refs.~ 
\cite{Colangelo:1992kc,Colangelo:1998ga,Dai:1996yw,Dai:1998ca}), where the separation of the two lowest $1^+$ states from each other is straightforward, due to their  different orbital angular momentum $j$. One has to
choose an appropriate interpolating current with $j=1/2$ or 
$j=3/2$ and consider two separate correlators. 
However, for a finite $c$-quark mass the two $1^+$ states mix
in the observed $D_1$ and $D_1'$ mesons, making 
the HQET sum rules inadequate for these mesons.
Correlators with a finite $c$-quark mass were used in, e.g.,  Refs.~\cite{Colangelo:1991ug,Eletsky:1995bz}.
Still, these calculations assume that only one $1^+$ lowest state is interpolated by the conventional axial current $\bar{c}\gamma_\mu\gamma_5q$. 
This is however in contradiction with HQET that predicts a second $1^+$
state in the same mass region, which has been confirmed by experimental data. Consequently, the assumption that only a single state is present within the duality window of a QCD sum rule cannot be justified.  

QCD light-cone sum rules (LCSRs) \cite{Balitsky:1986st,Balitsky:1989ry,Chernyak:1990ag} opened up
new possibilities to calculate the $B$-to-charm  form factors, especially their version with  
$B$-meson distribution amplitudes (DAs) proposed in Refs.~\cite{Khodjamirian:2005ea,Khodjamirian:2006st}.
These sum rules are derived for a finite $c$ quark mass, which makes them suitable for our task. They can be used in principle for any charmed hadron in the final state, 
 and not only for $D$ and $D^*$ mesons as  in, e.g., Refs.~\cite{Faller:2008tr,Wang:2017jow,Gubernari:2018wyi}.
Nevertheless, we encounter also in this approach the problem of defining a current that interpolates only one of the $1^+$ states at a time.

In this paper, we suggest a novel procedure to separate nearby resonances that supersedes the standard LCSR approach. 
This procedure consists in defining two independent currents 
with $J^P=1^+$, which in general interpolate both the $D_1$ and $D_1'$ mesons.
By finding a suitable linear combinations of these currents that interpolate only one  $1^+$ state at 
a time, we can write down the desired LCSRs.
However, these linear combinations of currents depend on four unknown decay constants (one for each  meson-current combination).
Using two-point QCD sum rules, we are able to determine only three out of the four decay constants.
The fourth one is determined a posteriori, by using the experimental measurement of the $ B\to D_1 \ell 
\nu_\ell$ decay width.
After determining this remaining unknown parameter, we predict the $ B\to D_1$ and $ B\to D_1'$ form factors.
\\
 
The rest of the paper is organized as follows. 
In \refsec{2ptSR} we derive the two-point sum rules for the decay constants and fix three out of the four decay constants that enter in our calculation.  
In \refsec{LCSR} we define the dedicated interpolating currents and derive the  LCSRs.
In \refsec{Nres} we fix the remaining unknown decay constant by fitting 
the expression for the $ B\to D_1 \ell 
\nu_\ell$ total decay width to its measured value.
Then, we predict the  $B\to D_1^{(\prime)}$ form factors, the $D_1^{(\prime)}$-mesons decay constants, and the LFU ratios $R(D_1^{(\prime)})$.
A series of appendices contain details regarding the $B$-meson DAs (in \ref{app:BDA}), our analytical results for the LCSRs (in \ref{app:OPEcoeff}), and the formulae for the $ B\to D_1^{(\prime)} \ell \nu_\ell$ differential decay widths (in \ref{app:width}).

\section{Two-point QCD sum rules for the $1^+$ charmed mesons}
\label{sec:2ptSR}

To derive the sum rules for the decay constants of the $1^+$ charmed mesons, we 
construct the two-point correlators
\begin{eqnarray} 
    \label{eq:corr2pt}
    \Pi_{\mu \nu }^{(ij)}  (q) = i \int d^4 x \, e^{iqx} \langle 0 | \mathcal{T}\{ J^{(i)}_\mu (x) J^{(j)\dagger}_\nu (0)   \} | 0 \rangle 
    =- g_{\mu\nu}\Pi^{(ij)}(q^2)  +   q_\mu q_\nu \widetilde{\Pi}^{(ij)} (q^2)
    \,,
\end{eqnarray} 
for $i,j=1,2$. 
The currents in \refeq{corr2pt} are defined as
\begin{align}
\label{eq:J1}
    J^{(1)}_\mu & = (m_c + m_q) \bar{c} \gamma_\mu \gamma_5  q 
    \,,
    \\* 
    J^{(2)}_\mu & = 
    i \bar{c} \gamma_5 \overleftrightarrow{D}_\mu  q
    \,,\label{eq:J2}
\end{align}
where $\overleftrightarrow{D}_\mu \equiv D_\mu - \overleftarrow{D}_\mu\,$ and $q=u,\,d$.
Throughout this paper we work in the isospin limit and do not 
distinguish the flavors of the two light quarks, whose masses are neglected.

The two currents (\ref{eq:J1})-(\ref{eq:J2}) are linearly independent since they interpolate different states:
While the current $J^{(1)}_\mu$ creates from the vacuum (apart from $0^-$ states)  $1^+$ states, 
for which the light degrees of freedom have exclusively an angular momentum of  $j = 1/2$, the current $J^{(2)}_\mu$  interpolates, in addition, $1^+$ states with    
 $j_{\rm} = 3/2$. In \refeq{J1} we also added a factor $(m_c + m_q)$ to the 
dimension-3 operator, such that both currents have the same mass dimension. 
\\

The hadronic states with spin-parity $J^P=1^+$ contribute to both invariant amplitudes in Eq.~(\ref{eq:corr2pt}), since their contributions
are proportional to the transverse combination 
$( q^2g_{\mu\nu}+q_\mu q_\nu\!)$.
Conversely, the pseudoscalar states 
--- which start from $D$ meson --- contribute only to the amplitude multiplying the $q_{\mu}q_{\nu}$ 
structure.\footnote{ 
    The situation similar to 
    the QCD sum rule for the light axial meson first obtained in 
    Ref.~\cite{Shifman:1978bx}. The diagonal sum rules for (conventional) 
    heavy-light axial currents were considered in Ref.~\cite{Eletsky:1995bz}. } 
Hence, we consider
the invariant amplitude $\Pi^{(ij)}(q^2)$ to isolate the axial charmed meson contributions.

We also notice that the currents $J^{(1)}_\mu$ and $J^{(2)}_\mu$ interpolate both the $D_1$ and the $D_1'$ mesons. 
Therefore, to extract the decay constants of these mesons, we need to consider simultaneously the three independent sum rules for the relevant correlators: $\Pi^{(11)}$, $\Pi^{(22)}$, and $\Pi^{(12)}$ since $\Pi^{(12)}=\Pi^{(21)}$.
We express these correlators in terms of hadronic dispersion relations in \refsec{hadrepr}, while we  compute the same correlators using an operator product expansion (OPE) in \refsec{OPEcal}.
These two representations of the correlators are then matched.
In addition, semi-global quark-hadron duality is used to remove the contribution of the continuum and further excited states.
A Borel transform is then performed to reduce the systematic uncertainty due to quark-hadron duality. 
The procedure up to this point is a standard one \cite{Shifman:1978bx} with many successful applications in the literature (see, e.g., the review \cite{Colangelo:2000dp}).
In this article, we extend this procedure to deal with the case where there are two mesons with very close masses in the same spin-parity channel, like the $D_1$ and the $D_1'$ mesons.

\subsection{Hadronic representation of the two-point correlators}
\label{sec:hadrepr}

The hadronic spectrum of the correlators defined in \refeq{corr2pt} --- considering only the $J^P=1^+$ channel --- consists of the two low-lying resonances $D_1$ and $D'_1$ and the continuum with further excited states.
This spectrum is markedly
different from the one  in a ``typical'' QCD sum rule with a single ground-state resonance.  
The currently available experimental data
suggest that  the well established $D_1(2420)$ resonance is narrow, whereas  the $D'_1(2430)$ resonance
is very broad and  has a mass only about 10 MeV smaller (see \reftab{spectr}). 
Therefore, the hadronic dispersion relation for the three correlators
(\ref{eq:corr2pt}) after performing the Borel transform reads
\begin{align}
    \Pi_{\HAD}^{(11)}(M^2)
    & =
    f_{D_1}^2 m_{D_1}^4 
    e^{-m_{D_1}^2/M^2} 
    + 
    f_{D_1'}^2 m_{D_1'}^4 
    e^{-m_{D_1'}^2/M^2}
    +
    \int\limits_{s_{\th}}^\infty ds\,e^{-s/M^2}\rho_{\cont}^{(11)}(s)
    \,,
    \label{eq:HAD11}
    \\
    \Pi_{\HAD}^{(12)}(M^2)
    & =
    f_{D_1}g_{D_1} m_{D_1}^4 
    e^{-m_{D_1}^2/M^2} 
    + 
    f_{D_1'}g_{D_1'} m_{D_1'}^4 
    e^{-m_{D_1'}^2/M^2}
    +
    \int\limits_{s_{\th}}^\infty ds\,e^{-s/M^2}\rho_{\cont}^{(12)}(s)
    \,,
    \label{eq:HAD12}
    \\
    \Pi_{\HAD}^{(22)}(M^2)
    & =
    g_{D_1}^2 m_{D_1}^4 
    e^{-m_{D_1}^2/M^2} 
    + 
    g_{D_1'}^2 m_{D_1'}^4 
    e^{-m_{D_1'}^2/M^2}
    +
    \int\limits_{s_{\th}}^\infty ds\,e^{-s/M^2}\rho_{\cont}^{(22)}(s)
    \,,
    \label{eq:HAD22}
\end{align}
where $s_{\th} = (m_{D} + 2m_\pi)^2\simeq (m_{D^*} + m_\pi)^2$ is the lowest 
threshold of hadronic continuum states in this channel  and $M^2$ is the Borel parameter.
The decay constants introduced in the above equations are defined as
\begin{equation}
\begin{aligned}
    \langle 0 | J^{(1)}_\mu |  D_1(q,\lambda) \rangle  = m_{D_1}^2 \varepsilon_\mu^{(D_1)} f_{D_1}\,,~~~
   \langle 0 | J^{(1)}_\mu |  D'_1(q,\lambda) \rangle = m_{D_1}^2 \varepsilon_\mu^{(D'_1)} f_{D'_1}\,,
   \\
   \langle 0 | J^{(2)}_\mu |  D_1(q,\lambda) \rangle = m_{D_1}^2 \varepsilon_\mu^{(D_1)} g_{D_1}\,,~~~
   \langle 0 | J^{(2)}_\mu |  D'_1(q,\lambda) \rangle = m_{D_1}^2 \varepsilon_\mu^{(D'_1)} g_{D'_1}\,,
\label{eq:fgD1}
\end{aligned}
\end{equation}
where $\varepsilon_\mu^{(D_1^{(\prime)})}\equiv \varepsilon_\mu^{(D_1^{(\prime)})}(p,\lambda)$ is the polarization vector of the $D_1^{(\prime)}$ mesons.
In \refeqs{HAD11}{HAD22}, we have isolated the ground-state resonances and 
attributed a generic hadronic spectral density 
$\rho_{\cont}^{(ij)}(s)$ 
to the rest of the hadronic spectrum including continuum and excited states.

To make this 
hadronic representations more accurate, we  take into account
the large total width of the $D_1'$ meson, replacing
 the zero-width resonance by 
a Breit-Wigner form with the energy-dependent width. 
After performing the Borel transform, this corresponds to 
the replacement of a simple exponential in \refeqs{HAD11}{HAD22}
by the following expression: 
\begin{eqnarray}
e^{-m_{D_1'}^2/M^2}\to \E(\Gamma_{D_1'},M^2)=\int\limits^{\infty}_{s_{\th}}\!ds\, e^{-s/M^2}\Bigg[
\frac{1}{\pi}
\frac{\sqrt{s}\Gamma_{D_1'}(s)}{(s-m_{D_1'}^2)^2+s\Gamma^2_{D_1'}(s)}\Bigg]
\,.
\label{eq:widthfact}
\end{eqnarray}
The formula for the energy dependence of $\Gamma_{D_1'}(s)$ is approximated
by the $S$-wave 
phase-space factor for the dominant decay channel $D_1'\to D^*\pi$ with the largest phase space:
\begin{eqnarray}
\Gamma_{D_1'}(s)=\Gamma_{D_1'}^{\rm tot}
\Bigg[\frac{\lambda^{1/2}(s,m_{D^*}^2,m_\pi^2)m_{D_1'}}{
\lambda^{1/2}(m_{D_1'}^2,m_{D^*}^2,m_\pi^2)\sqrt{s}}\Bigg]
\,,
\end{eqnarray}
where $\lambda$ is the K\"allen function.
In the narrow-width limit, that is for $\Gamma_{D_1'}^{\rm tot}\to 0$, it is easy to show that $\E(\Gamma_{D_1'},M^2) = e^{-m_{D_1'}^2/M^2}$.
The ansatz (\ref{eq:widthfact}) can be interpreted as a result of the resummation of 
the $D^*\pi$ intermediate states strongly coupled to the $D_1'$ resonance.
In other words, we effectively take into account the most important $D^*\pi$ continuum state with the lowest threshold
in the hadronic spectrum of the correlator.\footnote{
    This interpretation of the energy-dependent width
    for a broad $\rho$ resonance is explained, e.g., in
    Ref.~\cite{Bruch:2004py}, while for a similar  resonance ansatz in the 
    LCSRs for $B\to 2\pi$ form factors see Ref.~\cite{Cheng:2017smj}.}
Since the spectral density of this state spans up to $s\to \infty$,
in the final sum rules the upper limit 
of integration over $s$ in \refeq{widthfact} is replaced by the effective threshold of the quark-hadron duality.
This means that the part of the $D^*\pi$-state contribution above that threshold is subtracted as a part of the duality approximation.
We do not apply the replacement (\ref{eq:widthfact}) for the $D_1$ meson, since its width is relatively small and hence the narrow-width limit is a reasonable approximation in this case.

Finally, \refeqs{HAD11}{HAD22} become
\begin{align}
    \Pi_{\HAD}^{(11)}(M^2)
    & =
    f_{D_1}^2 m_{D_1}^4 
    e^{-m_{D_1}^2/M^2} 
    + 
    f_{D_1'}^2 m_{D_1'}^4 
    \E(\Gamma_{D_1'},M^2)
    +
    \int\limits_{s_{\th}}^\infty ds\,e^{-s/M^2}\rho_{\cont}^{(11)}(s)
    \,,
    \label{eq:HAD11BW}
    \\
    \Pi_{\HAD}^{(12)}(M^2)
    & =
    f_{D_1}g_{D_1} m_{D_1}^4 
    e^{-m_{D_1}^2/M^2} 
    + 
    f_{D_1'}g_{D_1'} m_{D_1'}^4 
    \E(\Gamma_{D_1'},M^2)
    +
    \int\limits_{s_{\th}}^\infty ds\,e^{-s/M^2}\rho_{\cont}^{(12)}(s)
    \,,
    \label{eq:HAD12BW}
    \\
    \Pi_{\HAD}^{(22)}(M^2)
    & =
    g_{D_1}^2 m_{D_1}^4 
    e^{-m_{D_1}^2/M^2} 
    + 
    g_{D_1'}^2 m_{D_1'}^4 
    \E(\Gamma_{D_1'},M^2)
    +
    \int\limits_{s_{\th}}^\infty ds\,e^{-s/M^2}\rho_{\cont}^{(22)}(s)
    \,.
    \label{eq:HAD22BW}
\end{align}

\subsection{OPE of the two-point correlators}
\label{sec:OPEcal}

We compute the correlators \eqref{eq:corr2pt} by expanding the time-ordered product for $x \sim 0$, that is using a short-distance OPE. 
The leading power contribution to this OPE is reduced to  the perturbative calculation of the correlator. 
The higher-power terms --- which are typically suppressed by inverse powers of $M^2$ --- are given by a series of QCD vacuum condensates multiplied by the corresponding Wilson coefficient.
As a result, the correlators can be decomposed as
\begin{align}
    \Pi_{\OPE}^{(ij)}(q^2) = \Pi_{\pert}^{(ij)}(q^2) +  \Pi_{\cond}^{(ij)}(q^2)
    \,.
\end{align}
We discuss the calculation of $\Pi_{\pert}^{(ij)}$ and $\Pi_{\cond}^{(ij)}$ in the remainder of this section. 
\\

The spectral densities of the correlators $\Pi_{\pert}^{(ij)}$, defined as $\rho^{(ij)}(s)\equiv (1/\pi)\Im \Pi^{(ij)}(s) $, 
at leading order (LO) read
\bea
 \rho_{\pert}^{(11)}(s)&=&
 \frac{m_c^2\left(s-m_c^2\right)^2 \left(m_c^2+2 s\right)}{8 \pi ^2 s^2}
 \theta\left(s-m_c^2\right)
   \,, 
   \label{eq:rhopert11}
   \\
\rho_{\pert}^{(12)}(s) &=& 
- \frac{m_c^2 \left(s-m_c^2\right)^3}{8 \pi ^2 s^2}\theta\left(s-m_c^2\right)
   \,,
  \label{eq:rhopert12} 
 \\
 \rho_{\pert}^{(22)}(s)&=&
 \frac{\left(s-m_c^2\right)^4}{8 \pi ^2 s^2}\theta\left(s-m_c^2\right)
 \,.
\label{eq:rhopert22}
\eea
The next-to-leading order (NLO) gluon radiative correction to $\Pi^{(11)}_\pert$ is known 
(see, e.g., Ref.~\cite{Eletsky:1995bz}).
This correction is numerically not large, 
in contrast to the case of $b$-flavored currents \cite{Gelhausen:2014jea}.
To obtain the same correction for  the correlators containing the current $J_\mu^{(2)}$ 
one would have to perform
a dedicated calculation of two-loop diagrams, which is out of the scope of this work.
Hence, for consistency, we do not include the NLO corrections in the OPE. 
\\

The QCD vacuum condensates are organized in series with increasing mass dimension of the respective operators.
We consider contributions up to and including $d=5$ condensates:
\begin{align}
    \Pi_{\cond}^{(ij)} =
    \Pi_{\bar qq}^{(ij)} +
    \Pi_{GG}^{(ij)} +
    \Pi_{\bar qGq}^{(ij)}
    \,.
\end{align}
Here $\Pi_{\bar qq}^{(ij)}$, $\Pi_{GG}^{(ij)}$, and $\Pi_{\bar qGq}^{(ij)}$ are the quark, gluon, and quark-gluon condensate contributions, respectively.

To calculate the $d=3$ quark condensate terms we use the vacuum average 
$$
\langle 0 | \bar q_{\alpha i} q^k_\beta | 0\rangle = \frac{1}{4N_c} \langle  \bar q q  \rangle \delta_i^k \delta_{\alpha\beta}\,.
$$
We find 
\begin{align}
\label{eq:condqq}
&
\Pi_{\bar qq}^{(11)}(q^2) = \langle \bar  q q \rangle \frac{m_c^3}{m_c^2 - q^2} \,,
& &
\Pi_{\bar qq}^{(12)}(q^2) = 
\Pi_{\bar qq}^{(22)}(q^2) =  0\,.
& 
\end{align}
Note that including the next-to-leading term in the expansion of  quark field at $x=0$, namely
\bea
\label{eq:qexpan}
\bar{q}(x)=\bar{q}(0)+\left.x^{\mu} \bar{q}(x) \overleftarrow{D}_{\mu}\right|_{x=0}+\left.\frac{1}{2} x^{\mu} x^{\nu} \bar{q}(x) \overleftarrow{D}_{\mu} \overleftarrow{D}_{\nu}\right|_{x=0}+\ldots\,,
\eea
generates contributions in Eq.~\eqref{eq:condqq} proportional to the light-quark mass and hence they can be neglected. 

The $d=4$ contributions to Eq.~(\ref{eq:condqq}) proportional to the gluon condensate density
\bea
\langle G G \rangle \equiv \frac{\alpha_s}{\pi} \bra{0} G_{\mu\nu}^a G^{\mu\nu a}\ket{0}
\nonumber
\eea
are conveniently calculated adopting
 the Fock-Schwinger gauge, defined as $$(x-x_0)^\mu A^a_\mu(x)=0 $$ for $x_0=0$  (see, e.g., the reviews \cite{Novikov:1983gd,Khodjamirian:2020btr}).
\begin{figure}[t]
\centerline{	\includegraphics[scale=1.05]{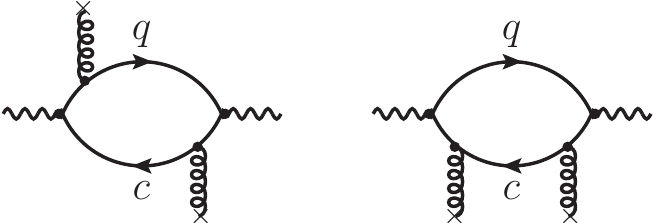}}
	\caption{\emph{The diagrams with nonvanishing contributions 
	to the gluon condensate term in the OPE of the two-point correlators. The crossed lines denote vacuum gluons. }}
	\label{fig:diagGG}  
\end{figure} 
In Fig.~\ref{fig:diagGG} we display the diagrams 
which provide non vanishing contributions to the gluon condensate term in the OPE. The remaining 
diagram where the light quark emits two gluons vanishes in the adopted limit $m_q=0$. 
Also the additional diagrams for $\Pi^{(12),(22)}$ with a gluon emitted from the
vertex $J^{(2)}_\mu$ with covariant derivative are vanishing.
The  diagrams in Fig.~\ref{fig:diagGG} yield
\begin{align}
    \Pi_{GG}^{(11)}(q^2) & = -\frac{1}{12}\langle G G \rangle \frac{m_c^2}{m_c^2-q^2}\,,
    \label{eq:GG11}
    \\
    \Pi_{GG}^{(12)}(q^2) & = 
    \frac{1}{24 (q^2)^2} \langle G G \rangle m_c^2
    \left(\left(m_c^2-3 q^2\right) \log \left(\frac{m_c^2}{m_c^2-q^2}\right)-q^2\right)
    \,,
    \label{eq:GG12}
    \\
    \Pi_{GG}^{(22)}(q^2) & =  \frac{1}{24 (q^2)^2}\langle G G \rangle \bigg(-m_c^2 q^2 
    +\left(-4 m_c^2
    q^2+m_c^4+3 (q^2)^2\right) \log \left(\frac{m_c^2}{m_c^2-q^2}\right)\bigg)
    \,.
    \label{eq:GG22}
\end{align}
In the last equation we omit the constant terms since they vanish after performing the Borel transform. 
The functions in \refeqs{GG12}{GG22} that multiply
the gluon condensate densities develop  imaginary parts
at $q^2=s>m_c^2$:
\begin{align}
    \rho_{GG}^{(12)}(s)
    & =
    \frac{  m_c^2
   \left(m_c^2-3  s\right) }{24 s^2}\theta \left(s-m_c^2\right)\,,
   \\
    \rho_{GG}^{(22)}(s)
    & =
    \frac{  \left(m_c^4-4 m_c^2 s+3
   s^2\right)}{24 s^2}\theta \left(s-m_c^2\right)\,.
\label{eq:shoGG}
\end{align}
Therefore, the gluon condensate contributions (\ref{eq:GG12}) and (\ref{eq:GG22}) 
can be  represented in the form of an unsubtracted dispersion relation:
\begin{equation}
\Pi_{GG}^{(12),(22)}(q^2)=\left\langle GG \right\rangle\int\limits_{m_c^2}^{\infty}\!\frac{ds}{s-q^2}\,\rho_{GG}^{(12),(22)}(s)\,.
\label{eq:GGdisp}
\end{equation}
We treat these contributions in the same way as the perturbative contributions
of \refeqs{rhopert11}{rhopert22}, i.e. we include  them into the OPE
spectral density.

Truncating  the short-distance OPE at $d=5$,
we also take into account  the quark-gluon condensate:
\bea
\langle\bar{q} G q\rangle  \equiv \langle 0 |\bar{q} g_{s} G_{\mu \nu}^{a} t^{a}\sigma^{\mu \nu} q | 0 \rangle
= m_0^2 \langle \bar q q \rangle
\,.
\nonumber
\eea
The calculation of this contribution --- although technically more complicated than for the quark
condensate --- is well documented in the reviews such as Refs.~\cite{Novikov:1983gd,Khodjamirian:2020btr} and  hence we do not dwell on details. 
We only mention that here again the diagrams with a gluon emitted from the vertex containing a covariant derivative vanish. 
The results for the quark-gluon
condensate contributions are
\begin{equation}
\begin{aligned}
\label{eq:condqGq}
&
\Pi_{\bar qGq}^{(11)}(q^2) =  - \frac{1}{2} m_0^2 \langle \bar q q \rangle \frac{m_c^5}{(m_c^2-q^2)^3}\,,
& &
\Pi_{\bar qGq}^{(22)}(q^2) = \frac{3}{8} m_0^2 \langle \bar q q \rangle   \frac{m_c}{(m_c^2-q^2)} \,, \\ &
\Pi_{\bar qGq}^{(12)}(q^2) = 
\frac{1}{12} m_0^2 \langle \bar q q \rangle   \frac{m_c\, q^2}{(m_c^2-q^2)^2}\,.
& 
\end{aligned}
\end{equation}
Finally, we note that neglecting the $d=6$ four-quark condensate contributions 
is justified because in the correlators of heavy-light currents these corrections
are in general numerically negligible (for example in the sum rule for the $D^*$ decay constant of Ref.~\cite{Gelhausen:2014jea}).

Following the standard procedure to derive a sum rule, we perform a Borel transform of the results listed in this section.
We obtain the following expressions for the OPE of the three 
correlators to the adopted accuracy:
\begin{align}
\Pi_\OPE^{(11)}(M^2)&=\int\limits_{m_c^2}^\infty\!ds \,e^{-s/M^2}\rho^{(11)}_{\pert}(s)+
\Big[\langle \bar{q}q\rangle m_c^3-\frac{1}{12} \langle GG\rangle m_c^2
-\frac{1}{4}m_0^2\langle \bar{q}q\rangle \frac{m_c^5}{M^4}\Big]e^{-m_c^2/M^2}
\,,
\label{eq:PiM11}
\\
\Pi_\OPE^{(12)}(M^2)&=
\int\limits_{m_c^2}^\infty\!ds \,e^{-s/M^2}\Big[\rho^{(12)}_{\pert}(s)+
\langle GG\rangle \rho^{(12)}_{GG}(s)\Big]
-\frac{1}{12}m_0^2\langle \bar{q}q\rangle m_c\left(1-\frac{m_c^2}{M^2}\right)e^{-m_c^2/M^2}
\,,
\label{eq:PiM12}
\\
\Pi_\OPE^{(22)}(M^2)&=\int\limits_{m_c^2}^\infty\!ds \,e^{-s/M^2}\Big[\rho^{(22)}_{\pert}(s)+
\langle GG\rangle \rho^{(22)}_{GG}(s)\Big]
+\frac{3}{8}m_0^2\langle \bar{q}q\rangle m_c e^{-m_c^2/M^2}
\,.
\label{eq:PiM22}
\end{align}

\subsection{Two-point sum rules and upper bounds}
\label{sec:num}

To obtain the two-point sum rules for the decay constants, we match the hadronic representation of the correlators 
given in \refsec{hadrepr} to their respective OPE calculations given in \refsec{OPEcal}.
We also use semi-global quark-hadron duality to remove
the contributions of the continuum and excited states encoded in the functions $\rho_{\cont}^{(ij)}$.
This implies that the upper limit  of the integrals in the OPE results (\ref{eq:PiM11})-(\ref{eq:PiM22}) and, simultaneously, the upper limit  in the expression for $\E(\Gamma_{D_1'},M^2)$ are replaced by the respective effective thresholds $s_0^{(ij)}$, whose choice is discussed in \refsec{Nres}.
The resulting sum rules for the three two-point correlators read
\begin{align}
    f_{D_1}^2 m_{D_1}^4 
    e^{-m_{D_1}^2/M^2} 
    + 
    f_{D_1'}^2 m_{D_1'}^4 
    \E(\Gamma_{D_1'},M^2,s_0^{(11)})
    & =
    \Pi_\OPE^{(11)}(M^2,s_0^{(11)})
    \,,
    \label{eq:sum11}
    \\
    f_{D_1}g_{D_1} m_{D_1}^4 
    e^{-m_{D_1}^2/M^2} 
    + 
    f_{D_1'}g_{D_1'} m_{D_1'}^4 
    \E(\Gamma_{D_1'},M^2,s_0^{(12)})
    & =
    \Pi_\OPE^{(12)}(M^2,s_0^{(12)})
    \,,
    \label{eq:sum12}
    \\
    g_{D_1}^2 m_{D_1}^4 
    e^{-m_{D_1}^2/M^2} 
    + 
    g_{D_1'}^2 m_{D_1'}^4 
    \E(\Gamma_{D_1'},M^2,s_0^{(22)})
    & =
    \Pi_\OPE^{(22)}(M^2,s_0^{(22)})
    \,.
    \label{eq:sum22}
\end{align}
Since there are only three independent sum rules for four unknown decay constants, it is not possible to determine all of them.
We then consider \refeqs{sum11}{sum22} as a system of equations to express $f_{D_1'}$, $g_{D_1}$, and $g_{D_1'}$  as a function of $f_{D_1}$.
This system has four distinct solutions, which are discussed in detail in \refsec{Nres}. 
These decay constants serve as input parameters for the LCSRs, which are derived in the next section.
\\

Even though we cannot calculate the individual decay constants with the information inferred from the sum rules, we can still set bounds on their value.
In fact, the two 
diagonal correlators $\Pi^{(11)}_{\mu\nu}$ and $\Pi^{(22)}_{\mu\nu}$ have positive
definite spectral densities. 
Hence, independent of the quark-hadron duality assumption, 
the following  model-independent upper bounds are valid 
for the squared decay constants
(see, e.g., Ref.~\cite{Khodjamirian:2008xt} where the upper bounds for the $D_{(s)}$ decay constants
were obtained):
\begin{equation}
\begin{aligned}
&
f_{D_1}^2 < \frac{\Pi_{\OPE}^{(11)}(M^2)}{m_{D_1}^4 e^{-m_{D_1}^2/M^2}}\, ,
&\qquad&
f_{D_1'}^2 < \frac{\Pi_{\OPE}^{(11)}(M^2)}{m_{D_1'}^4 {\cal E}(\Gamma_{D_1'},M^2)}\,,
&
\\
&
g_{D_1}^2 < \frac{\Pi_{\OPE}^{(22)}(M^2)}{m_{D_1}^4 e^{-m_{D_1}^2/M^2}}\, ,
&&
g_{D_1'}^2 < \frac{\Pi_{\OPE}^{(22)}(M^2)}{m_{D_1'}^4 {\cal E}(\Gamma_{D_1'},M^2)}\,.
&
\label{eq:uplim}
\end{aligned}
\end{equation}
The numerical value of these bounds are presented in \refsec{Nres}.

\section{LCSRs for the $B\to D_1^{(\prime)}$ form factors}
\label{sec:LCSR}

Here we derive the  LCSRs with $B$-meson  DAs for the  $B\to D_1$ and $B\to D_1'$ form factors.
This method is well established and has already been applied to several  $B$-meson transitions.
For the derivation of our sum rules we follow Refs.~\cite{Faller:2008tr,Gubernari:2018wyi}.

In analogy with the two-point sum rules considered in the previous section, we start 
by defining suitable $B$-to-vacuum correlators:
\begin{align}
\label{eq:corrF}
&
{\cal F}^{(R)}_{\mu\nu}(p, q) = i \!\int \!d^4 x\, e^{ip\cdot x}\, \langle 0|
\T \left\{ J^{(R)\dagger}_\nu(x), J^{\rm w}_\mu(0) \right\} | \bar{B}(p+q)\rangle\,,
&&
(R=D_1,\,D_1')\,.
&
\end{align}
Here, $J^{\rm w}_\mu=\bar{c}\gamma_\mu(1-\gamma_5)b$ is the weak current with the momentum $q$
while $p+q$ is the momentum of the $B$ meson state, so that $(p+q)^2=m_B^2$.
The interpolating currents 
$J_\mu^{(D_1)}$ or $J_\mu^{(D_1')}$ with momentum $p$ are chosen such that they only interpolate the $D_1$ or $D_1'$ meson, respectively. 
This can be easily achieved by combining the decay constants defined in \refeq{fgD1}:
\begin{equation}
J_\mu^{(D_1)}=J_\mu^{(1)}- \frac{f_{D_1'}}{g_{D_1'}}J_\mu^{(2)}, ~~~~~ J_\mu^{(D_1')}=J_\mu^{(1)}
-\frac{f_{D_1}}{g_{D_1}} J_\mu^{(2)}\,.
\label{eq:mixJ}
\end{equation}

Again, the correlators (\ref{eq:corrF}) can be both expressed in terms of hadronic quantities and computed using --- in this case --- a light-cone OPE.
The details of these calculations are given in \refsec{LCSRhad} and \refsec{LCSROPE}, respectively.
The LCSRs are then obtained by matching the results of the hadronic representation and OPE calculation and using semi-global quark-hadron duality.

\subsection{Hadronic representation of the \textbf{\textit{B}}-to-vacuum correlator} 
\label{sec:LCSRhad}

To obtain the hadronic dispersion relation for the correlators (\ref{eq:corrF}), we calculate the imaginary part with respect to the variable $p^2$, by inserting a complete set of hadronic states in \refeq{corrF}:
\begin{equation}
\begin{aligned}
    {\rm Im}_{p^2} {\cal F}^{(R)}_{\HAD,\mu\nu}(p, q)
    &=
    \frac12 
    \sum \!\!\!\!\!\!\!\! \int \limits_{h}
    d\tau_h
    (2\pi)^4 \delta^{(4)}(p_h - p)
    \langle 0|J^{(R)\dagger}_\nu|\overline{h(p)\rangle
    \langle h(p)} | J^{\rm w}_\mu | \bar{B}(p+q)\rangle
    \,,
\label{eq:ImcorrF}
\end{aligned}
\end{equation}
where the overline denotes the  sum over  different polarizations of
a given intermediate state $h$.
The contributions of the $D_1$ and $D_1'$ mesons to \refeq{ImcorrF} read
\begin{equation}
\begin{aligned}
    {\rm Im}_{p^2} {\cal F}^{(R)}_{\HAD,\mu\nu}(p, q)
    &=
    \pi
    \,\delta(s - m_R^2)
    \langle 0|J^{(R)\dagger}_\nu|\overline{R(p)\rangle
    \langle R(p)} | J^{\rm w}_\mu | \bar{B}(p+q)\rangle
    +\dots
    \,,
    \label{eq:1pthad}
\end{aligned}
\end{equation}
where the ellipsis stands for contributions from other hadronic states with the same
quantum numbers.
Using the definitions in \refeq{fgD1}, one can easily find that
\begin{equation}
\begin{aligned}
    \langle0|J^{(D_1)}_\nu|D_1(p)\rangle
    & =
    m_{D_1}^2 \varepsilon_\nu^{(D_1)}
    \left(
    f_{D_1} -
    \frac{f_{D_1'}}{g_{D_1'}}
    g_{D_1}
    \right)
    \,,\\
    \langle0|J^{(D_1')}_\nu|D_1'(p)\rangle
    & =
    m_{D_1'}^2 \varepsilon_\nu^{(D_1')}
    \left(
    f_{D_1'} -
    \frac{f_{D_1}}{g_{D_1}}
    g_{D_1'}
    \right)
    \,,
    \label{eq:decCon}
\end{aligned}
\end{equation}
and, by construction, $$\langle0|J^{(D_1')}_\nu|D_1(p)\rangle = 
\langle0|J^{(D_1)}_\nu|D_1'(p)\rangle=0\,.$$
Furthermore, we express the hadronic 
matrix elements of the weak current in terms of the $B\to D_1^{(\prime)}$ form factors:
\begin{multline}
\langle R(p,\varepsilon)|J^{\rm w}_\mu | \bar{B}(p+q)\rangle 
= -i
\varepsilon^{(R)*}_\mu (m_B+m_{R})V_{1}^{BR}(q^2)
+
i (2p+q)_\mu (\varepsilon^{(R)*}\cdot q) \frac{V_{2}^{BR}(q^2)}{m_B+m_{R}}
\\*
+ i q_\mu(\varepsilon^{(R)*}\cdot q)
\frac{2 m_{R}}{q^2}
\left(
    V_{3}^{BR}(q^2) - V_{0}^{BR}(q^2)
\right)
-\epsilon_{\mu\nu\alpha\beta}\varepsilon^{(R)*\nu} p^\alpha q^\beta \frac{2 A^{BR}(q^2)}{m_B+m_{R}}
\,,
\label{eq:BD1FF}
\end{multline}
where $
2 m_{R} V_{3}^{BR}(q^2) = 
(m_B + m_{R}) V_{1}^{BR}(q^2)
-(m_B - m_{R}) V_{2}^{BR}(q^2)
$ and $V_{0}^{BR}(0) = V_{3}^{BR}(0)$\,,
and we adopt the convention $\epsilon^{0123}=1$.
Note that our form factor definitions are analogous to the conventional definitions of the $B\to D^*$ form factors of, e.g., Ref.~\cite{Faller:2008tr}. 
There are  now one axial-vector and three independent vector form factors instead of one vector and three axial-vector form factors in the $B\to D^*$ case, since the parity of the final state is opposite.

Substituting the definitions of hadronic matrix elements (\ref{eq:decCon})-(\ref{eq:BD1FF}) in \refeq{1pthad} we obtain
\begin{align}
    \frac{1}{\pi}{\rm Im}_{p^2} {\cal F}^{(D_1)}_{\HAD,\mu\nu}(p, q)
    &= m_{D_1}^2
    h_{D_1} 
    \Bigg[ 
    i\left(g_{\mu\nu}-\frac{p_\mu p_\nu}{m_{D_1}^2}\right) (m_B+m_{D_1})V_{1}^{BD_1}(q^2)
    \nonumber\\*
    &- i(2p+q)_\mu\left( q_\nu-p_\nu\frac{(q\cdot p)}{m_{D_1}^2}\right) \frac{V_{2}^{BD_1}(q^2)}{m_B+m_{D_1}}
    \nonumber\\*
    &- i\left( q_\mu q_\nu-q_\mu p_\nu \frac{(q\cdot p)}{m_{D_1}^2}\right) 
    \frac{2 m_{D_1}}{q^2}\left(V_{3}^{BD_1}(q^2) - V_{0}^{BD_1}(q^2)\right)
    \nonumber\\*
    &+ \epsilon_{\mu\nu\alpha\beta}p^\alpha q^\beta \frac{2A^{BD_1}(q^2)}{m_B+m_{D_1}}
    \Bigg]\delta(p^2-m_{D_1}^2)
    +\dots\,,
\label{eq:D1contrib}
    \\
    \frac{1}{\pi}{\rm Im}_{p^2} {\cal F}^{(D_1')}_{\HAD,\mu\nu}(p, q)
    &= m_{D_1'}^2
    h_{D_1'} 
    \Bigg[ 
    i\left(g_{\mu\nu}-\frac{p_\mu p_\nu}{m_{D_1'}^2}\right) (m_B+m_{D_1'})V_{1}^{BD_1'}(q^2)
    \nonumber\\*
    &- i(2p+q)_\mu\left( q_\nu-p_\nu\frac{(q\cdot p)}{m_{D_1'}^2}\right) \frac{V_{2}^{BD_1'}(q^2)}{m_B+m_{D_1'}}
    \nonumber\\*
    &- i\left( q_\mu q_\nu-q_\mu p_\nu \frac{(q\cdot p)}{m_{D_1'}^2}\right) 
    \frac{2 m_{D_1'}}{q^2}\left(V_{3}^{BD_1'}(q^2) - V_{0}^{BD_1'}(q^2)\right)
    \nonumber\\*
    &+ \epsilon_{\mu\nu\alpha\beta}p^\alpha q^\beta \frac{2A^{BD_1'}(q^2)}{m_B+m_{D_1'}}
    \Bigg]
\frac{1}{\pi}
\frac{\sqrt{p^2}\Gamma_{D_1'}(p^2)}{(p^2-m_{D_1'}^2)^2+p^2\Gamma^2_{D_1'}(p^2)}
+\dots\,.
\label{eq:D1pcontrib}
\end{align}
Here, as in \refsec{hadrepr}, we take into account the width of the $D_1'$ state and, for the sake of brevity, introduce a notation for the two combinations of decay constants:
$$
h_{D_1}\equiv \left(f_{D_1} -\frac{f_{D_1'}}{g_{D_1'}}g_{D_1}\right), ~~~
h_{D_1'}\equiv \left(f_{D_1'} -\frac{f_{D_1}}{g_{D_1}}g_{D_1'}\right)\,.
$$

For future convenience, we decompose the correlator ${\cal F}^{(R)}_{\mu\nu}(p, q)$ in a set of independent Lorentz structures ${\cal L}_{\mu\nu}(p, q)$ multiplied by the corresponding invariant amplitudes ${\cal F}_{{\cal L}}^{(R)}(p^2,q^2)$:
\bea
{\cal F}^{(R)}_{\mu\nu}(p, q)=\sum_{\cal L} {\cal L}_{\mu\nu}(p,q) {\cal F}_{\cal L}^{(R)}(p^2,q^2)\,.
\label{eq:LexpcorrF}
\eea
From the above decomposition we specifically choose the Lorentz structures 
\begin{equation}
{\cal L}_{\mu\nu}=g_{\mu\nu},~ p_\mu q_\nu ,~q_\mu q_\nu\,,~\epsilon_{\mu\nu\alpha\beta}p^\alpha q^\beta\,,
\label{eq:Lstr}
\end{equation}
since they are free from the contributions of pseudoscalar charmed mesons. 
Indeed, the hadronic 
matrix elements $\langle 0|J^{(R)\dagger}_\nu|D(p)\rangle$ are proportional to the momentum $p_\nu$, which  does not appear in \refeq{Lstr}.
Finally, we can write down a hadronic dispersion relation for each function ${\cal F}_{\cal L}^{(R)}(p^2,q^2)$:
\begin{align}
    {\cal F}_{\HAD,\L}^{(R)}(p^2,q^2)=\frac{1}{\pi} \int\limits_{s_{\th}}^\infty \!ds 
    \,\frac{{\rm Im}_{p^2} {\cal F}_{\HAD,\L}^{(R)}(s,q^2)}{s-p^2}\,,
\label{eq:dispF}
\end{align}
where $s_{\th}$ is the same threshold  as in the hadronic dispersion relations 
for the two-point correlator considered  in \refsec{hadrepr}.

\subsection{Light-cone OPE of the \textbf{\textit{B}}-to-vacuum correlator}
\label{sec:LCSROPE}

The calculation of the light-cone OPE for the correlators (\ref{eq:corrF}) is analogous to the one performed in  Refs.~\cite{Khodjamirian:2006st,Faller:2008tr,Gubernari:2018wyi} for the $B\to P$ or $B\to V$ form factors,
where $P=\pi,K,D$ and $V=\rho,K^*, D^*$.
The main difference with these calculations is the choice of the interpolating current, which has already been discussed at the beginning of this section. 

It is more convenient to consider separately the correlators
\begin{align}
\label{eq:LCcorri}
&
{\cal F}^{(i)}_{\mu\nu}(p, q) = i \!\int \!d^4 x\, e^{ip\cdot x}\, \langle 0|
\T \left\{ J^{(i)\dagger}_\nu(x), J^{\rm w}_\mu(0) \right\} | \bar{B}(p+q)\rangle\,,
&&
(i=1,\,2)
&
\end{align}
which are related to the ones in \refeq{corrF} through the following equations:
\begin{align}
     \label{eq:FiFR}
     &
     {\cal F}^{(D_1)}_{\mu\nu}
     = {\cal F}^{(1)}_{\mu\nu}
     - \frac{f_{D_1'}}{g_{D_1'}} {\cal F}^{(2)}_{\mu\nu}
     \,,
     &&
     {\cal F}^{(D_1')}_{\mu\nu}
     = {\cal F}^{(1)}_{\mu\nu}
     - \frac{f_{D_1}}{g_{D_1}} {\cal F}^{(2)}_{\mu\nu}
     \,.
     &
\end{align}
We expand the correlators ${\cal F}^{(i)}_{\mu\nu}$ at near light-cone separations $x^2 \simeq 0$, which in momentum space implies that our calculation is valid for $p^2\ll m_c^2$ and $q^2 \ll (m_b+m_c)^2$.
Using the Wick's theorem to compute the time-ordered product in \refeq{LCcorri} and expanding the $c$-quark propagator near the light-cone, we obtain
\begin{align}
\label{eq:corrF11}
    {\cal F}^{(1)}_{\OPE,\mu\nu}(p, q) 
    & = 
    i\!\int \!d^4 x\, e^{ip\cdot x}\, \langle 0|
    m_c\bar{q}(x)\gamma_\nu\gamma_5
    i S_c(x,0)
    \Gamma_\mu^\text{w} b(0) | \bar{B}(p+q)\rangle\,,
    \\
    \label{eq:corrF2}
    {\cal F}^{(2)}_{\OPE,\mu\nu}(p, q) 
    & = 
    \!\int \!d^4 x\, e^{ip\cdot x}\, \langle 0|
    \bar{q}(x)\gamma_5(D_\nu-\overleftarrow{D}_\nu)
    i S_c(x,0)
    \Gamma_\mu^\text{w} b(0) | \bar{B}(p+q)\rangle\,,
\end{align}
where $\Gamma_\mu^\text{w} =  \gamma_\mu(1-\gamma_5)$.
We neglect the gluon emission effects which generate quark-antiquark-gluon  
(three-particle) components of the $B$ meson DAs.
Indeed, their  
contributions  turned out to be numerically irrelevant in 
LCSRs for the $B\to D^{(*)}$ form factors ~\cite{Gubernari:2018wyi,Gubernari:2020eft}.
We also do not take into account the $O(\alpha_s)$ corrections  which can be relevant ~\cite{Wang:2017jow}
but would demand involved loop diagram calculations which are out of our scope.
Thus, the covariant derivatives in Eq.~(\ref{eq:corrF2}) can be replaced with  partial derivatives and  in both correlation functions the free $c$-quark propagator 
can be used:
\bea
\label{eq:Scprop1}
S_c(x,0)= \int\frac{d^4k}{(2\pi)^4}\,e^{-ik\cdot x}\frac{ \slashed{k} +m_c}{k^2-m_c^2}\,.
\eea
Using these approximations, \refeqs{corrF11}{corrF2} can be written as
\begin{align}
\label{eq:corrF12}
    {\cal F}^{(1)}_{\OPE,\mu\nu}(p, q) 
    & = 
    - m_c \!\int \!d^4 x\, e^{ip\cdot x}\, 
    \langle 0|\bar{q}^\alpha(x)b^\beta(0)  | \bar{B}(p+q)\rangle
    \big[\gamma_\nu \gamma_5 S_c(x,0)\Gamma_\mu^\text{w} \big]_{\alpha\beta} \,,
    \\
    \label{eq:corrF3}
    {\cal F}^{(2)}_{\OPE,\mu\nu}(p, q) 
    & = 
    i \!\int \!d^4 x\, e^{ip\cdot x}\, 
    \Big\{ 
    \langle 0|\bar{q}^\alpha(x)b^\beta(0)  | \bar{B}(p+q)\rangle
    \big[\gamma_5\partial_\nu S_c(x,0)\Gamma_\mu^\text{w} \big]_{\alpha\beta}
    \nonumber\\*
    & - 
    \langle 0|\bar{q}^\alpha(x)\overleftarrow{\partial}_\nu b^\beta(0)  | \bar{B}(p+q)\rangle
    \big[\gamma_5 S_c(x,0)\Gamma_\mu^\text{w} \big]_{\alpha\beta}\Big\} \,,
\end{align}
where $\alpha,\beta$ are Dirac indices.
\refeq{corrF3} is further simplified integrating by parts its second line: 
\begin{align}
\label{eq:corrF2compact}
{\cal F}^{(2)}_{\OPE,\mu\nu}(p, q) 
& =  
i \!\int \!d^4 x\, e^{ip\cdot x}\, 
 \langle 0|\bar{q}^\alpha(x)b^\beta(0)  | \bar{B}(p+q)\rangle
\nonumber\\*
& \times
 \Big\{
2\big[\gamma_5 \partial_\nu S_c(x,0)\Gamma_\mu^\text{w} \big]_{\alpha\beta}
+ip_\nu
\big[\gamma_5 S_c(x,0)\Gamma_\mu^\text{w} \big]_{\alpha\beta}\Big\}\,.
\end{align}
This allows us to write the correlators ${\cal F}^{(1)}_{\OPE,\mu\nu}$ and ${\cal F}^{(2)}_{\OPE,\mu\nu}$ in the same compact form
\begin{align}
\label{eq:corrFgen}
{\cal F}^{(i)}_{\OPE,\mu\nu}(p, q) = \!\int \!d^4 x\, e^{ip\cdot x}\!\!\int \frac{d^4k}{(2\pi)^4}e^{-ik\cdot x}
\big[\Gamma^{(i)}_\nu \gamma_5
\frac{\slashed{k}+m_c}{m_c^2-k^2}
\Gamma_\mu^\text{w} \big]_{\alpha\beta}
\langle 0|\bar{q}^\alpha(x)b^\beta(0)  | \bar{B}(p+q)\rangle
\,,    
\end{align}
where 
\begin{align*}
    &\Gamma^{(1)}_\nu=m_c\gamma_\nu\,,& &\Gamma^{(2)}_\nu=p_\nu-2k_\nu\,.&
\end{align*}

To proceed, we approximate the $B$-to-vacuum  matrix element in the above expression by its HQET limit,  replacing the
$b$-quark field by the heavy-quark effective field 
with velocity $v=(p+q)/m_B$, i.e. $b(0)\to h_v(0)$. 
This non-local $B$-to-vacuum  matrix element can now be expanded in $B$-meson light-cone DAs of increasing twist.
We emphasize that even though we apply the HQET approximation for the $b$ quark and light 
degrees of freedom within $B$ meson, we still treat the virtual  
$c$ quark  in the correlation functions as a full QCD object with a finite mass.

In our calculation of correlation functions we specifically use the DAs given in Ref.~\cite{Braun:2017liq} up to twist four. In addition, we include the twist-five DA $g_-$ in the Wandzura-Wilczek limit as in Ref.~\cite{Gubernari:2018wyi}.
For the reader's convenience we collect the relevant formulae in \refapp{BDA}.
Using the expressions given there, taking the traces, and isolating the Lorentz structures $\L_{\mu\nu}$ listed  in (\ref{eq:Lstr}), \refeq{corrFgen} can be written as
\begin{align}
\label{eq:OPEres}
    {\cal F}^{(i)}_{\OPE,\L}(p^2, q^2)
    =
    f_B m_B\sum_k \int\limits_0^\infty d\sigma 
    \frac{
        I_\L^{(i,k)}(\sigma,q^2)
    }{
        \left( p^2 -s \right)^k
    }
    \,, ~~~~ (i=1,2)   
\end{align}
where
\begin{align}
    &
    \sigma\equiv\frac{\omega}{m_B} \,,
    &&
    s(\sigma,q^2)\equiv\sigma m_B^2+
    \frac{m_c^2-\sigma q^2}{\bar{\sigma}} \,,
    &&
    \bar \sigma
    \equiv 1-\sigma \,.
    &
\end{align}
The functions $I_\L^{(i,k)}$ are linear combinations of the four $B$-meson DAs:
\bea
I_\L^{(i,k)}(\sigma,q^2)=
\sum_\psi \C^{(i,k)}_{\L,\psi} (\sigma,q^2)\psi(m_B\sigma)\,,
\label{eq:Ink}
\eea
for $\psi=\phi_+,g_+,\overline{\Phi}_{\pm}, \overline{G}_{\pm}$.
Our results for the coefficients $\C^{(i,k)}_{\L,\psi}$
are collected in \refapp{OPEcoeff}.
This completes our OPE calculation for the correlators ${\cal F}^{(1)}_{\mu\nu}$ and ${\cal F}^{(2)}_{\mu\nu}$.

\subsection{Light-cone sum rules}

To obtain the LCSRs for the $B\to D_1^{(\prime)}$ form factors we match the hadronic 
representations of the correlators from  \refsec{LCSRhad} onto their respective OPE expressions presented in \refsec{LCSROPE}. 
In addition, we use semi-global quark-hadron duality to eliminate the contribution of the continuum and excited states in the hadronic dispersion relation.
After adopting this approximation and performing the Borel transform, the OPE result (\ref{eq:OPEres}) can be written as~\cite{Gubernari:2018wyi,Descotes-Genon:2019bud}
\begin{align}
    {\cal F}^{(i)}_{\OPE,\L}(\hat{M}^2, \hat{s}_0, q^2)
    &=
    f_B m_B \sum_{k=1}^{4}
    \frac{(-1)^{k}}{(k-1)!}
    \Bigg\{\int_{0}^{\sigma_0} d \sigma \;e^{-s(\sigma,q^2)/\hat{M}^2} \frac{1}{(\hat{M}^2)^{k-1}}
    I_\L^{(i,k)}(\sigma,q^2)
    \nonumber\\*
    & + \Bigg[e^{-s(\sigma,q^2)/\hat{M}^2}\sum_{j=1}^{k-1}\frac{1}{(\hat{M}^2)^{k-j-1}}\frac{1}{s'}
    \left(\frac{d}{d\sigma}\frac{1}{s'}\right)^{j-1}
    I_\L^{(i,k)}(\sigma,q^2)
    \Bigg]_{\sigma=\sigma_0}\Bigg\rbrace\,,~~ (i=1,2)\,,
    \label{eq:MasterFor1}
\end{align}
where $\hat{s}_0$ is the LCSRs effective threshold.  We assume 
a universal duality interval for all LCSRs.
In the equation above, we have introduced the following notation:
\begin{align*}
&
    \left(\frac{d}{d\sigma}\frac{1}{s'}\right)^{n} f(\sigma)
    \equiv
    \left(\frac{d}{d\sigma}\frac{1}{s'}\left(\frac{d}{d\sigma}\frac{1}{s'}\dots f(\sigma)\right)\right) \,,
    &&
    s' \equiv \frac{ds}{d\sigma} \,,
    &
    \\
    &
    \sigma(q^2,s)=
    \frac{
    m_B^2-q^2+s-\sqrt{4 \left(m_c^2-s\right)
    m_B^2+\left(m_B^2-q^2+s\right)^2}
    }{
    2 m_B^2
    }\,,
    &&
    \sigma_0 \equiv \sigma(q^2,\hat{s}_0)
    \,.
    &
\end{align*}
Finally, we can write down the LCSRs
$B\to D_1$ and $B\to D_1'$ form factors: 
\begin{align}
\,m_{D_1}^2
h_{D_1}
\frac{2 A^{BD_1}(q^2)}{m_B+m_{D_1}}e^{-m_{D_1}^2/\hat{M}^2}
&=
{\cal F}^{(D_1)}_{\OPE,\epsilon_{\mu\nu pq}}\!(\hat{M}^2, \hat{s}_0,q^2)
\,,
\label{eq:AlcsrB1}
\\
i\, m_{D_1}^2
h_{D_1}
(m_B+m_{D_1})V_1^{BD_1}(q^2)e^{-m_{D_1}^2/\hat{M}^2}
&=
{\cal F}^{(D_1)}_{\OPE,g_{\mu\nu}}(\hat{M}^2, \hat{s}_0, q^2)
\,,
\label{eq:V1lcsrB1}
\\
-i\,2m_{D_1}^2
h_{D_1}
\frac{V_2^{BD_1}(q^2)}{m_B+m_{D_1}}e^{-m_{D_1}^2/\hat{M}^2}
&=
{\cal F}^{(D_1)}_{\OPE,p_\mu q_\nu}(\hat{M}^2, \hat{s}_0, q^2)
\,,
\label{eq:V2lcsrB1}
\\
-2i\,m_{D_1}^3 
h_{D_1}
\frac{
    V_{3}^{BD_1}(q^2) - V_{0}^{BD_1}(q^2)}
    {q^2}
e^{-m_{D_1}^2/\hat{M}^2}
&=
{\cal F}^{(D_1)}_{\OPE,r_\mu q_\nu}(\hat{M}^2, \hat{s}_0, q^2) 
\label{eq:V30lcsrB1}
\,, \\
\,m_{D_1'}^2
h_{D_1'}
\frac{2 A^{BD_1'}(q^2)}{m_B+m_{D_1'}}
\E(\Gamma_{D_1'},\hat{M}^2,\hat{s}_0)
&=
{\cal F}^{(D_1')}_{\OPE,\epsilon_{\mu\nu pq}}\!(\hat{M}^2, \hat{s}_0, q^2)
\,,
\label{eq:AlcsrB1p}
\\
i\, m_{D_1'}^2
h_{D_1'}
(m_B+m_{D_1'})V_1^{BD_1'}(q^2)
\E(\Gamma_{D_1'},\hat{M}^2,\hat{s}_0)
&=
{\cal F}^{(D_1')}_{\OPE,g_{\mu\nu}}(\hat{M}^2, \hat{s}_0, q^2)
\,,
\label{eq:V1lcsrB1p}
\\
-i\,2m_{D_1'}^2
h_{D_1'}
\frac{V_2^{BD_1'}(q^2)}{m_B+m_{D_1'}}
\E(\Gamma_{D_1'},\hat{M}^2,\hat{s}_0)
&=
{\cal F}^{(D_1')}_{\OPE,p_\mu q_\nu}(\hat{M}^2, \hat{s}_0, q^2)
\,,
\label{eq:V2lcsrB1p}
\\
-2i\,m_{D_1'}^3 
h_{D_1'}
\frac{
    V_{3}^{BD_1'}(q^2) - V_{0}^{BD_1'}(q^2)}
    {q^2}
\E(\Gamma_{D_1'},\hat{M}^2,\hat{s}_0)
&=
{\cal F}^{(D_1')}_{\OPE,r_\mu q_\nu}(\hat{M}^2, \hat{s}_0, q^2) 
\,.
\label{eq:V30lcsrB1p}
\end{align}
Here, the equations (\ref{eq:widthfact}),  (\ref{eq:D1contrib})-(\ref{eq:Lstr}), and (\ref{eq:FiFR})  have been used.
We have  also  introduced the shorthand notation
\begin{align}
     &
     {\cal F}^{(D_1)}_{\OPE,{\cal L}}
     = {\cal F}^{(1)}_{\OPE,{\cal L}}
     - \frac{f_{D_1'}}{g_{D_1'}} {\cal F}^{(2)}_{\OPE,{\cal L}}
     \,,
     &&
     {\cal F}^{(D_1')}_{\OPE,{\cal L}}
     = {\cal F}^{(1)}_{\OPE,{\cal L}}
     - \frac{f_{D_1}}{g_{D_1}} {\cal F}^{(2)}_{\OPE,{\cal L}}
     \,,
     &
\end{align}
and
\begin{align}
    {\cal F}^{(R)}_{\OPE,r_\mu q_\nu}
    \equiv
    {\cal F}^{(R)}_{\OPE,q_\mu q_\nu}
    - 
    \frac{1}{2}{\cal F}^{(R)}_{\OPE,p_\mu q_\nu}
    \,, ~~~ (R=D_1,D_1')\,.
\end{align}
Note that to extract the form factors $A^{BR}$, $V_1^{BR}$, and $V_2^{BR}$ we have selected the Lorentz structures $\epsilon_{\mu\nu\alpha\beta}p^\alpha q^\beta$, $g_{\mu\nu}$, and $p_\mu q_\nu$, respectively.
The form factor difference $V_{3}^{BR} - V_{0}^{BR}$ is extracted by taking the  linear combination of the Lorentz structures $q_\mu q_\nu$ and $p_\mu q_\nu$.

\section{Numerical analysis and predictions}
\label{sec:Nres}

Turning to the numerical analysis, we note
that each of the LCSRs for a $B \to D_1^{(\prime)}$ form factor 
presented in \refeqs{AlcsrB1}{V30lcsrB1p}
depends on all  four decay constants of $D_1^{(\prime)}$-mesons. 
In a standard LCSR with $B$-meson DAs, a form factor is 
multiplied by a single decay constant of the final-state hadron. This decay constant can be replaced by its analytical
expression inferred from the two-point sum rule. 
However,  here we only have at our disposal three  two-point sum rules 
(\ref{eq:sum11})-(\ref{eq:sum22}). Hence, one of the decay constants, which we choose to be 
$f_{D_1}$,  remains a  free parameter to be fixed from an 
additional external input specified below. 
The two-point sum rules provide three relations that 
allow us to express the three 
decay constants $f_{D_1'},\,g_{D_1}$ and $g_{D_1'}$ as functions of $f_{D_1}$.  
In more details, we adopt the following procedure:

\begin{itemize}
    \item We set the input parameters and evaluate the OPE part of the two-point sum rules  in Eqs.~(\ref{eq:sum11})-(\ref{eq:sum22}).
    
    \item Employing these sum rules, we express the decay constants $f_{D_1'}$, $g_{D_1}$, and $g_{D_1'}$  as functions of $f_{D_1}$, i.e. $f_{D_1'} = f_{D_1'}(f_{D_1})$, $g_{D_1} = g_{D_1}(f_{D_1})$ and $g_{D_1'} = g_{D_1'}(f_{D_1})$.
    
    \item Specifying the input parameters (including the 
    parameters of $B$ meson DAs, the interval of Borel mass $\hat{M}$ and the threshold $\hat{s}_0$), we evaluate the OPE part of the LCSRs ${\cal F}^{(i)}_{\OPE,\L}$ 
    in Eq.~(\ref{eq:MasterFor1}). This is done for negative values of the momentum transfer squared, where the OPE in the adopted approximation is valid.
    Using conformal mapping and a $z$ expansion, 
    we extrapolate these results to positive $q^2$ values.
    
    \item The extrapolated OPE results, together with the hadron masses and expressions for the decay constants,  
    are substituted in the LCSRs (\ref{eq:AlcsrB1})-(\ref{eq:V30lcsrB1p}) yielding   the $B \to D_1^{(\prime)}$ form factors as  functions of $f_{D_1}$.

    \item Employing these expressions for  $B \to D_1$ form factors, we compute the $B \to D_1 \ell \nu$ decay width as function of $f_{D_1}$.
    Matching this computation with the corresponding experimental value, we determine
    the numerical value of $f_{D_1}$.
    
    \item Finally, using  $f_{D_1}$, extracted with the procedure outlined
    above  we evaluate the numerical values of all $B \to D_1^{(\prime)}$ form factors 
    and predict the lepton-flavour universality ratios $R(D_1^{(\prime)})$. As a byproduct, the values of other decay constants are calculated as well.
\end{itemize}

\subsection{Numerical analysis of the sum rules and $f_{D_1}$ determination}
\label{sec:numSR}

The input parameters that we use for the numerical analysis 
outlined above are listed in Table~\ref{tab:inputs2pt}. 
For the parameters determining the OPE such as the $c$-quark mass, the vacuum condensate values and 
the characteristics of $B$ meson DAs, we quote their sources. 
Our choice of the sum-rule specific parameters deserves separate comments.
For instance, for the normalization scale we use a typical interval established 
in the analyses of other correlation functions with a virtual $c$-quark. 
In these cases (see, e.g., Refs.~\cite{Gelhausen:2014jea, Khodjamirian:2020mlb}) not only the
LO, as here, but also the NLO gluon radiative corrections were taken into account.
Importantly, their numerical effect turned out to be mild, indicating a good convergence of perturbative series at these particular scales.

In addition, to evaluate the OPE part of  the two-point sum rules, that is the r.h.s. of the \refeqs{sum11}{sum22}, we need to choose a suitable  interval for the Borel parameter $M^2$ and to determine the effective thresholds $s_0^{(ij)}$. The Borel parameter has to be chosen such that both the  contributions of states above $D_1^{(\prime)}$ and higher-power terms in the OPE are sufficiently suppressed.
An indicator of the goodness of this interval is a mild variation of the sum rule result.
For the sum rules (\ref{eq:sum11})-(\ref{eq:sum22}) 
the requirements listed above are fulfilled for  the interval quoted in Table~\ref{tab:inputs2pt}. 
Moreover, we find that the same interval 
represents a reasonable choice for the Borel parameter $\hat{M}^2$ in  LCSRs as well.

Concerning the choice of the duality threshold, the commonly used
procedure  consists in taking the ratio between the derivative of a sum rule 
over $-1/M^2$ and the initial sum rule. For the charmed meson channel this 
procedure was used, for example, in the LCSRs for the 
$B\to D^{(*)}$ form factors.
Note that here, apart from the conventional axial interpolating current,  we deal with a nonstandard current with a derivative. It is therefore  
important to clarify if a correlator of these currents 
reveals a peculiar duality threshold. To find that out, we have considered
the sum rule for the correlator $\Pi^{(22)}$ and applied the differentiation procedure to establish 
the value of $s_0^{(22)}$. The resulting interval for which 
the masses of the $D_1$ and $D_1'$ mesons (neglecting their small difference) 
are reproduced, is displayed in Table~\ref{tab:inputs2pt}. It also turns out to be in the same ballpark 
as the thresholds for various two-point sum rules and LCSRs with $D^*$ mesons (see, e.g., \cite{Gelhausen:2014jea,Faller:2008tr,Gubernari:2018wyi, Khodjamirian:2020mlb}).
Guided by this affinity and by the fact that the correlators $\Pi^{(11)}$ and $\Pi^{(12)}$ yield similar values for the threshold, we simplify our numerical analysis adopting one and the same threshold 
for all remaining sum rules. 
\\

\begin{table}[t!]
\centering
\begin{tabular}{|c|c|c|}
\hline
&&\\[-2mm]
Parameter & Value/Interval & Ref.
\\[2mm]\hline\hline
&&\\[-2mm]
normalization scale  & $\mu=1.5$\, GeV ~ (1.3\,-\,2.5) GeV& 
\cite{Gelhausen:2014jea, Khodjamirian:2020mlb} \\[1mm]
\hline
&&\\[-2mm]
$c$-quark mass & $m_c(\mu=1.5 \GeV)=1.205\pm 0.035$ GeV & \cite{pdg}
\\[2mm]\hline
&&\\[-2mm]
quark condensate  & 
$\langle \bar q q \rangle(\mu=1.5 \mbox{GeV})=
-\left(0.278\pm 0.022\; \mbox{GeV} \right)^3$& \cite{Aoki:2021kgd}
\\[2mm]\hline
&&\\[-2mm]
Ratio $\langle \bar{q}Gq \rangle/\langle \bar{q}q \rangle$ &
$m_0^2=0.8\pm 0.2$ GeV$^2$ &\\
&&\cite{Ioffe:2002ee}\\[-2mm]
Gluon condensate  &$\langle  GG \rangle= 0.012^{+0.006}_{-0.012}\,\mbox{GeV}^4$ &\\[2mm]
\hline &&\\[-2mm]
$B$-meson decay constant  & 
$f_B=189.4 \pm 1.4$\,MeV  & \cite{Bazavov:2017lyh}
\\[2mm] \hline
&&\\[-2mm]
Parameters of the & $\lambda_B=0.460\pm 0.110$\,GeV & \cite{Braun:2003wx} \\[3mm]
$B$-meson DAs~\tablefootnote{For the parameter $\lambda_B$ we use the value obtained from a QCD sum rule in Ref.~\cite{Braun:2003wx}. This value has recently been confirmed in Ref.~\cite{Khodjamirian:2020hob}.
For the parameters $\lambda_E^2$ and $\lambda_H^2$ we use the results of Ref.\cite{Nishikawa:2011qk}, which are in agreement --- due to the large uncertainties --- with an independent recent calculation of Ref.~\cite{Rahimi:2020zzo}.}  & $\lambda_E^2=0.03\pm 0.02$\,GeV$^2$ & \\
&& \cite{Nishikawa:2011qk} \\[-2mm]
& $\lambda_H^2=0.06\pm 0.03$\,GeV$^2$ &  \\[2mm]
\hline
&&\\[-2mm]
Borel parameters &  $M^2=(2.5 - 3.5)\,\GeV^2$\,, ~~$\hat{M^2}=M^2$~~&\\[1mm]
\hline
&&\\[-2mm]
Duality thresholds & $s_0^{(22)} = (7.20 \pm 0.65) \GeV^2 $& \\[1mm]
& $\hat{s}_0 = s_0 = s_0^{(11)}=s_0^{(12)}=  s_0^{(22)}$. &\\[1mm]
\hline
\end{tabular}
\caption{\emph{ Input values used in the numerical analysis of the two-point sum rules and the LCSRs. 
}
}
\label{tab:inputs2pt}
\end{table}

Using the inputs in \reftab{inputs2pt}, we evaluate the OPE part 
of the two-point sum rules and LCSRs in  \refeqs{sum11}{sum22} and (\ref{eq:MasterFor1}), respectively.
Inspecting the light-cone OPE in the LCSRs, 
we observe that for $q^2<0$ there is a strong suppression of   the higher-twist contributions (of the DAs $g_\pm$). 
However, at $q^2\geq 0$ these contributions almost reach the level of the lower-twist ones (of the DAs $\phi_\pm$). To ensure a maximal predictivity of the OPE results, we follow the same strategy as in Ref~\cite{Gubernari:2018wyi} and calculate the OPE at negative values,
choosing the points $q^2=\{-20.0,-15.0,-10.0,-5.0\}\GeV^2$.

We then extrapolate these OPE results to the semileptonic region $0<q^2 <(m_B - m_{D^{(\prime)}_1})^2$ using a parametrization similar to the one proposed in Refs.~\cite{Bourrely:2008za}. Each function ${\cal F}^{(i)}_{\OPE,\L}$ is written as
\begin{align}
    {\cal F}^{(i)}_{\OPE,\L} (q^2)
    =
    \frac{1}{1 - \frac{q^2}{m_{J^P}^2}}
    \sum_{k=0}^K
    \alpha_\L^{(k)}
    \left[
        z(q^2)
        -
        z(0)
    \right]^k
    \,,
    \label{eq:zexpOPE}
\end{align}
where
\begin{align}
    \label{eq:zdef}
    z(q^2) = \frac{\sqrt{t_+-q^2} - \sqrt{t_+ - t_0^{\phantom{1}}}}{\sqrt{t_+-q^2} + \sqrt{t_+ - t_0^{\phantom{1}}}}
    \,,
\end{align}
with the parameters
\begin{align}
    & t_+ = (m_B + m_D)^2 \,,&
    & t_0 = (m_B + m_D)\left(\sqrt{m_B} - \sqrt{m_D}\right)^2 \,.&
\end{align}
The parametrization (\ref{eq:zexpOPE}) isolates the $B_c(J^P)$ resonances located below
the  thresholds of the continuum $b\bar{c}$ states. The masses $m_{J^P}$ of these resonances, listed in \reftab{resonances}, 
have been calculated in lattice QCD.
One can see from \refeqs{AlcsrB1}{V30lcsrB1p} that each Lorentz structure is related  to a certain form factor, which in turn has certain spin-parity quantum numbers of the 
$b\bar c$ states in the timelike region. 
For the function ${\cal F}^{(i)}_{\OPE,r_\mu q_\nu}$, which corresponds to the form factors with both the $0^+$ and the $1^-$ states, we take for simplicity the mass $m_{0^+}$. Since $q_{\max}^2 = (m_B - m_{D_1^{(\prime)}})^2 \ll m_{0^+}^2$ this assumption has no significant numerical effect.
\begin{table}[t]
    \centering
    \renewcommand{\arraystretch}{1.25}
    \begin{tabular}{|c|c|c|}
        \hline
        $J^P$ & Form factors                                       & $B_c(J^P)$ resonance masses [$\GeV$] \\
        \hline
        \hline
        $0^+$ & $V_0^{BR}$                            & 6.70       \\\hline
        $1^-$ & $V_1^{BR}$, $V_2^{BR}$   & 6.33      \\\hline
        $1^+$ & $A^{BR}$                                         & 6.74      \\\hline
    \end{tabular}
    \caption{\emph{The lowest-lying $B_c(J^P)$ resonances in the form factor parametrizations. The masses are taken from the Refs.~\cite{pdg,Dowdall:2012ab}.
    \label{tab:resonances}
        }   }
\end{table}
Note that we truncate the series in \refeq{zexpOPE} already at $K=1$.
Given the uncertainties of the LCSRs, a third coefficient in this expansion should not affect significantly the results of the extrapolation.
\\

Having the OPE results in the semileptonic region, we obtain the $B\to D_1^{(\prime)}$ form factors in this region as  functions of $f_{D_1}$ using \refeqs{AlcsrB1}{V30lcsrB1p}.
Then, we calculate the 
total semileptonic width 
$\Gamma(B\to D_1\ell \nu)\equiv \Gamma_{D_1}^\th$ as a function of $f_{D_1}$.
The formulas for the decay width
expressed via form factors are given in \refapp{width}.
We determine $f_{D_1}$ by fitting
$\Gamma_{D_1}^\th (f_{D_1})$ to its experimental value obtained  
using the corrected experimental measurement quoted in Ref.~\cite{Bernlochner:2016bci  }
\be
\label{eq:expBR}
\B(B^+\to \bar{D}^0_1\ell^+\bar{\nu}_\ell)=(0.67\pm 0.05)\cdot 10^{-2},~~
(\ell =e,\mu)
\ee
 and
$\tau_{B^\pm}= 1.638$ ps  \cite{pdg}.

Note that, as we have already mentioned in \refsec{num}, there are
four solutions expressing the decay constants $f_{D_1'}$, $g_{D_1}$, and $g_{D_1'}$  in terms of $f_{D_1}$. 
This is due to the fact that \refeq{sum11} and \refeq{sum22} are quadratic in the decay constants.
However, only two of these solutions lead to phenomenologically different results. 
The other two solutions can be related to the former two via a redefinition of the  form-factor phase
which is not  observable.
As a result, the two independent solutions that  the fit 
 to \refeq{expBR} yields are
\begin{align}
    \label{eq:sol1}
    \text{sol. 1:}\qquad 
    f_{D_1}  &= (60\pm 20) \MeV  
    \,, \\*
    \label{eq:sol2}
    \text{sol. 2:}\qquad 
    f_{D_1} &= (95 \pm  25) \MeV \,.
\end{align}
We stress that the difference between the two solutions is not only in the resulting value of $f_{D_1}$, but also in the functional dependence of the other decay constants and the form factors on $f_{D_1}$.
For each of these solutions we  obtain numerical results for the full set of $B \to D_1^{(\prime)}$ form factors and $D_1^{(\prime)}$ decay constants.
These results are presented 
in the next subsection.
\\

The twofold ambiguity that emerges in our approach could be resolved by using additional experimental data.
However, the form factors obtained for 
both the solutions, (\ref{eq:sol1}) and (\ref{eq:sol1}), predict similar intervals of the $B\to D'_1\ell\nu_\ell$ width 
which is not used in the fit (see \refsec{FFsresl}).
These intervals, within large errors, are in agreement with the measured value.
Hence, the currently achieved accuracy of both LCSRs and experimental widths is not sufficient to distinguish the two solutions.
Certain angular observables might be able to resolve this twofold ambiguity, due to their different dependence on the form factors in contrast to the total decay width.
Alternatively, further theoretical inputs could also simplify the extraction of the $B \to D_1^{(\prime)}$ form factors within out approach.
For instance, if one of the $D_1^{(\prime)}$ decay constants was known from lattice QCD, we would not need to perform any fit to extract $f_{D_1}$ and thus we would be able to compute the form factor with no ambiguity.

Note that a systematic comparison of  our results with the ones 
obtained in Refs.~\cite{Leibovich:1997em,Bernlochner:2016bci} in the framework of HQET
is not straightforward and demands additional studies which are beyond our scope.
The evident reason for that is that in HQET  the
$D_1$ and $D_1'$ resonances are assumed to be pure $j=3/2$ and $j=1/2$ states,
respectively, whereas our approach does not use this assignment. 
Still, comparing the slopes of differential distributions 
obtained in \cite{Bernlochner:2016bci} with our predictions, we observe a better 
agreement when our form factors are obtained using ``sol. 1''.

\subsection{Form factors, decays constants, and LFU ratios}
\label{sec:FFsresl}

\begin{figure}[p]
\vspace{-2cm}
    \centering
    \newcommand\ww{0.40}
    \begin{tabular}{cc}
        \includegraphics[width=\ww\textwidth]{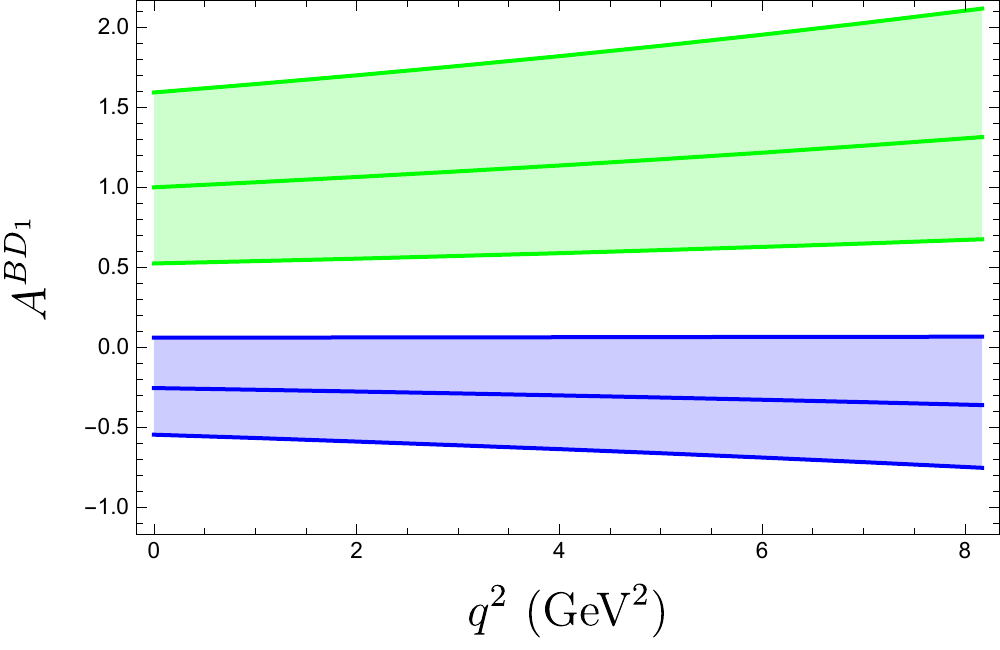}  &
        \includegraphics[width=\ww\textwidth]{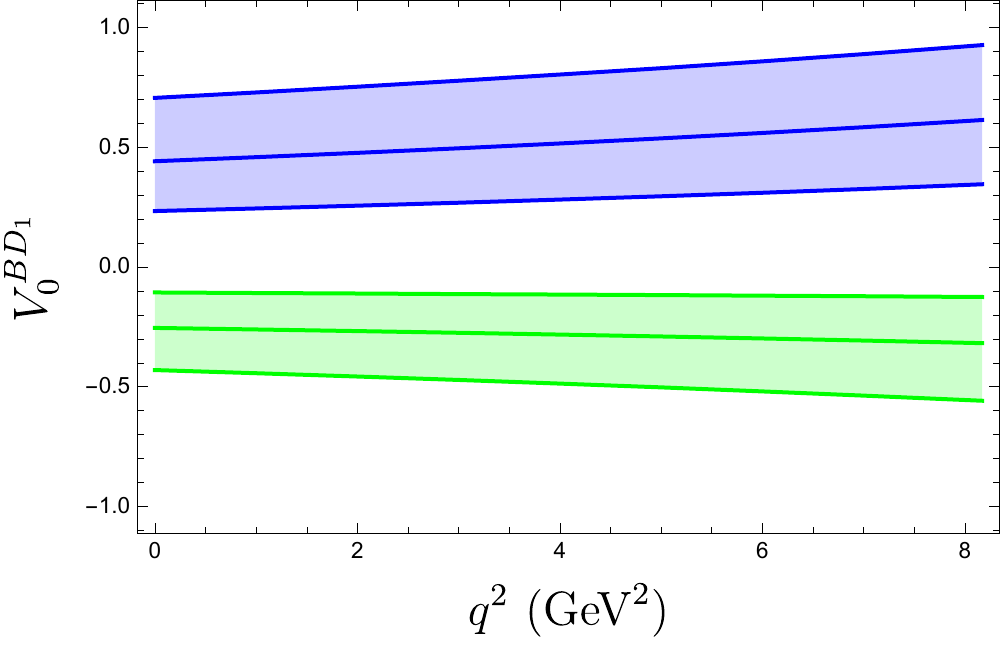}  \\[1.25em]
        \includegraphics[width=\ww\textwidth]{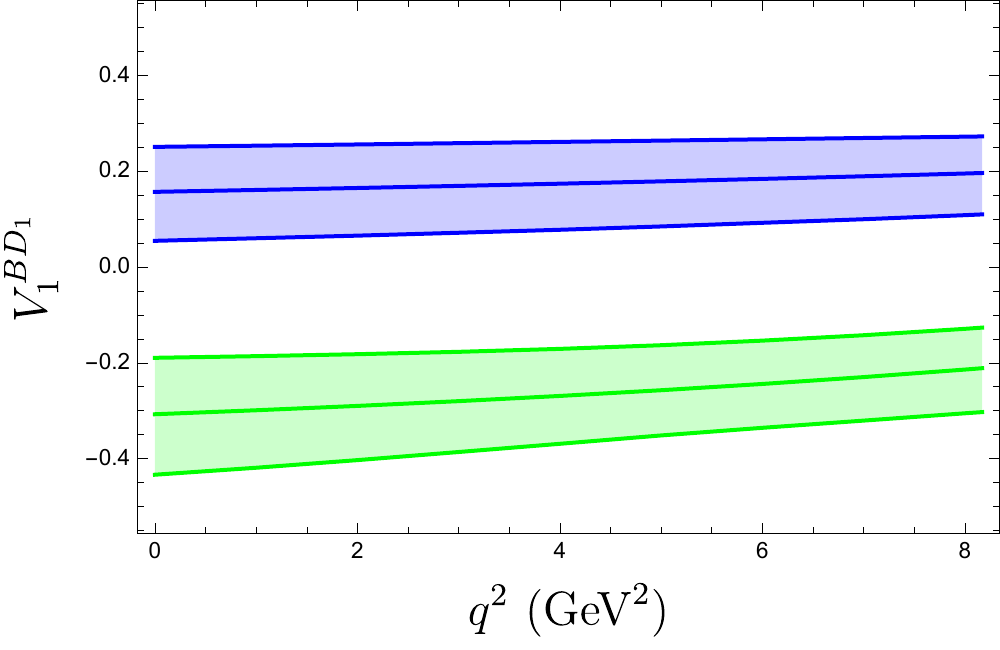}  &
        \includegraphics[width=\ww\textwidth]{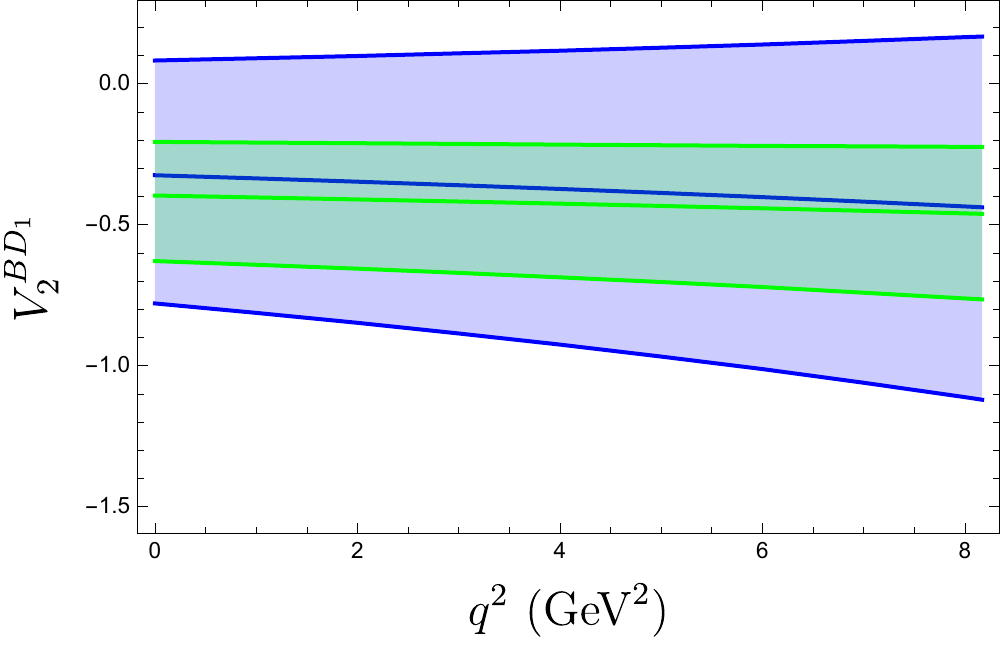}  \\[1.25em]
        \includegraphics[width=\ww\textwidth]{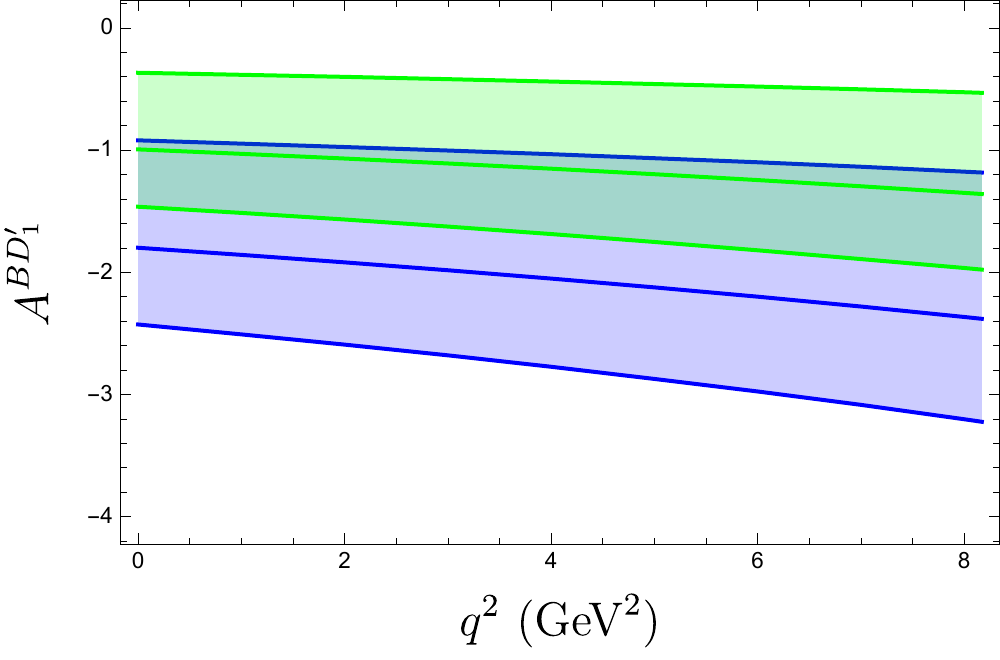}  &
        \includegraphics[width=\ww\textwidth]{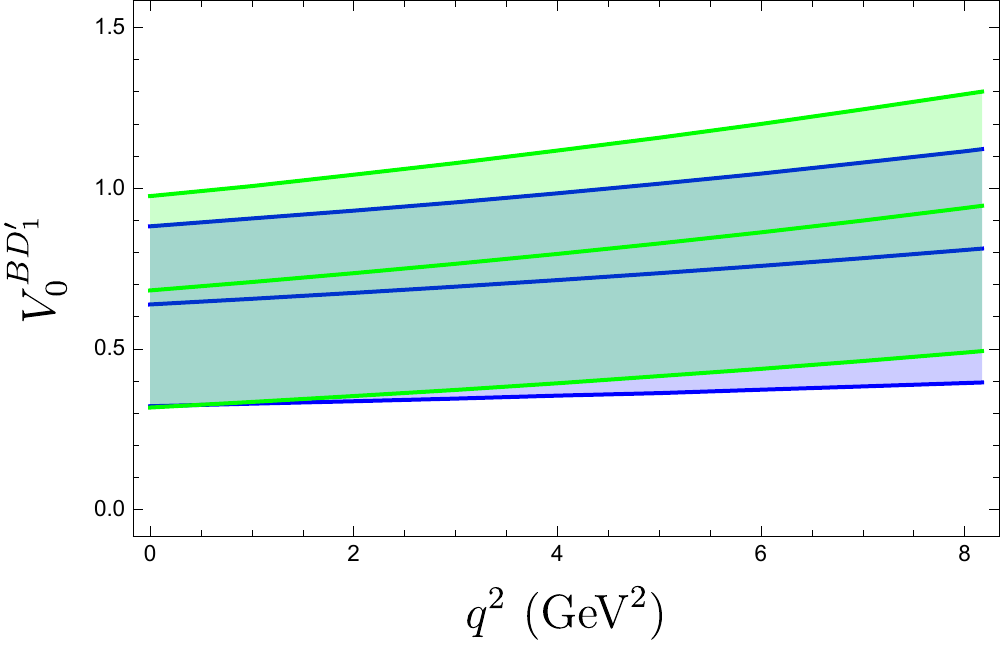}  \\[1.25em]
        \includegraphics[width=\ww\textwidth]{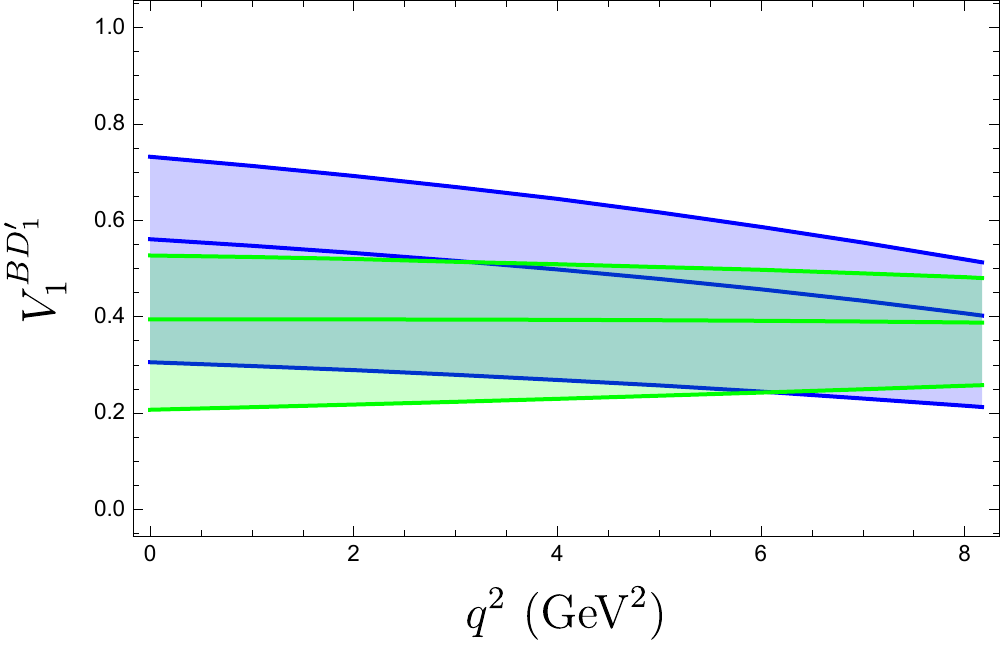}  &
        \includegraphics[width=\ww\textwidth]{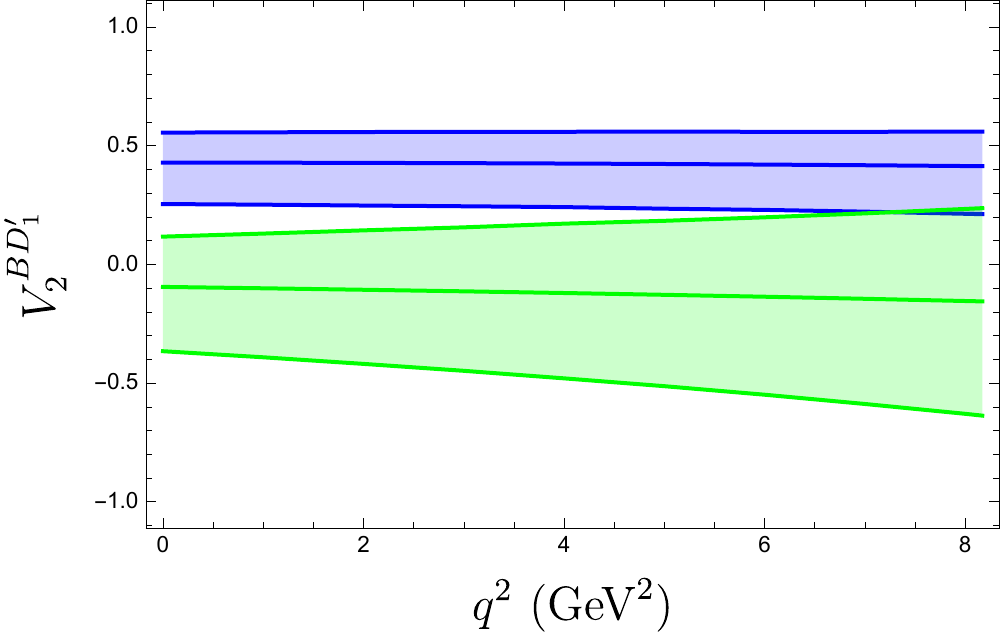}  \\[1.25em]
    \end{tabular}
    \caption{
    \emph{
    The $q^2$-dependence of the $B\to D_1$ and $B\to D_1'$
     form factors. The central values and 68\% probability envelopes in  blue (green) are 
    the results obtained using sol. 1 (sol. 2). 
    }}
    \label{fig:ffplots1}
\end{figure}

We predict the $B\to D_1$ and $B\to D_1'$ form factors using the OPE results obtained in  the previous subsection for the two alternative solutions emerging from our analysis.
The plots of these form factors are given in Fig.~\ref{fig:ffplots1}, where it is also  possible to observe the difference between
the two solutions in \refeqs{sol1}{sol2} of the fit to $f_{D_1}$.
For a practical use, we fit our form factor results 
to the following parametrization:
\begin{align}
    \label{eq:FFparam}
    &
    F^{BR} (q^2)
    =
    \frac{F^{BR} (0)}{1 - \frac{q^2}{m_{J^P}^2}}
    \left\{
    1
    +
    \beta_F
    \left[
        z(q^2)
        -
        z(0)
    \right]
    \right\}
    \,,
\end{align}
for $F^{BR}=A^{BR},V_0^{BR},V_1^{BR},V_2^{BR}$,
($R=D_1,D_1'$).
Here, we use the same  definition of the variable $z$ as in \refeq{zdef}.
The central values, the uncertainties, and the correlation of the coefficients in \refeq{FFparam} are given in \reftab{alphaF}.

\begin{table}[t]
    \centering
    \renewcommand{\arraystretch}{1.25}
    \begin{tabular}{|c|c|c|c|c|}
        \hline  
        &
        & 
        \hspace{1cm} $F^{BR} (0) $ \hspace{1cm}      &
        \hspace{1cm}  $\beta_F$ \hspace{1cm}         &
        Correlation                                  \\
        \hline\hline
\multirow{8}{*}{\rotatebox{90}{sol. 1}} &
$A^{BD_1}$ &
$ -0.27 \pm 0.29 $ &
$ -3.15 \pm 1.76 $ &
$ 0.03 $ \\
\cline{2-5}
        &
$V_0^{BD_1}$ &
$ 0.44 \pm 0.20 $ &
$ -3.41 \pm 1.26 $ &
$ 0.04 $ \\
\cline{2-5}
        &
$V_1^{BD_1}$ &
$ 0.16 \pm 0.10 $ &
$ 1.69 \pm 1.38 $ &
$ 0.01 $ \\
\cline{2-5}
        &
$V_2^{BD_1}$ &
$ -0.32 \pm 0.38 $ &
$ -4.19 \pm 6.29 $ &
$ 0.01 $ \\
\cline{2-5}
        &
$A^{BD_1'}$ &
$ -1.69 \pm 0.77 $ &
$ -1.82 \pm 0.74 $ &
$ 0.04$ \\
\cline{2-5}
        &
$V_0^{BD_1'}$ &
$ 0.60 \pm 0.32 $ &
$ -0.75 \pm 1.21 $ &
$ -0.04 $ \\
\cline{2-5}
        &
$V_1^{BD_1'}$ &
$ 0.53 \pm 0.22 $ &
$ 9.98 \pm 1.05 $ &
$ -0.02 $ \\
\cline{2-5}
        &
$V_2^{BD_1'}$ &
$ 0.40 \pm 0.15 $ &
$ 5.86 \pm 3.42 $ &
$ -0.02 $ \\
\hline\hline

\multirow{8}{*}{\rotatebox{90}{sol. 2}} &
$A^{BD_1}$ &
$ 1.00 \pm 0.45 $ &
$ -1.82 \pm 0.50 $ &
$ -0.06 $ \\
\cline{2-5}
        &
$V_0^{BD_1}$ &
$ -0.26 \pm 0.15 $ &
$ -0.26 \pm 1.63 $ &
$ 0.02 $ \\
\cline{2-5}
        &
$V_1^{BD_1}$ &
$ -0.31 \pm 0.12 $ &
$ 9.32 \pm 1.73 $ &
$ -0.02 $ \\
\cline{2-5}
        &
$V_2^{BD_1}$ &
$ -0.39 \pm 0.19 $ &
$ 2.26 \pm 2.36 $ &
$ 0.04 $ \\
\cline{2-5}
        &
$A^{BD_1'}$ &
$ -0.92 \pm 0.61 $ &
$ -3.28 \pm 1.26 $ &
$ -0.03$ \\
\cline{2-5}
        &
$V_0^{BD_1'}$ &
$ 0.66 \pm 0.33 $ &
$ -3.71 \pm 2.37 $ &
$ 0.03 $ \\
\cline{2-5}
        &
$V_1^{BD_1'}$ &
$ 0.37 \pm 0.19 $ &
$ 3.74 \pm 3.25 $ &
$ 0.01 $ \\
\cline{2-5}
        &
$V_2^{BD_1'}$ &
$ -0.12 \pm 0.25 $ &
$ -5.84 \pm 4.42 $ &
$ 0.02 $ \\
\hline

    \end{tabular}
    \caption{%
    \label{tab:alphaF}
    \emph{
    The central values, $1\sigma$ uncertainties, and correlations of the coefficients of the form factor parametrization (\ref{eq:FFparam}).
    }
    }
\end{table}

We predict also the decay constants $f_{D_1'}$, $g_{D_1}$, and $g_{D_1'}$ using the two-point sum rules derived in \refsec{2ptSR}. For the two solutions we obtain:
\begin{align}
    \text{sol. 1:}\qquad
    &
    f_{D_1'}  = -110\pm 28\,\MeV  \,, &
    &g_{D_1}   = -110\pm 38\,\MeV  \,, \quad &&
    g_{D_1'}  = -22\pm 31\,\MeV  \,.
    \\
    \text{sol. 2:}\qquad
    &
    f_{D_1'}  = 5\pm 34\,\MeV  \,, &&
    g_{D_1}   = 0\pm 34\,\MeV  \,, \quad &&
    g_{D_1'}  = -136\pm 44\,\MeV  \,.
\label{eq:decconst}
\end{align}
The predicted values for the decay constants
together with the fitted values (\ref{eq:sol1}),(\ref{eq:sol2})  for $f_{D_1}$, 
 lead to the following observation. 
 If the  solution~1  is adopted,  then the interpolating current $J^{(1)}_\mu$ has a 
larger overlap with the broad resonance $D_1'$ than with the narrow resonance $D_1$.
Simultaneously,  the current $J^{(2)}_\mu$ with derivative of quark field has a 
larger decay constant with $D_1$ than with $D_1'$. 
The solution 2 clearly manifests an opposite situation.
The observed correlation
between the structure of the interpolating currents and the 
decay constants  deserves further investigation. 

Furthermore, the values (\ref{eq:decconst}) together with the ones  for $f_{D_1}$ in  
\refeqs{sol1}{sol2} satisfy the upper bounds given in \refsec{num}.
Evaluating the r.h.s. of the inequalities (\ref{eq:uplim}) we obtain
\begin{equation}
\begin{aligned}
&
|f_{D_1}| < 181\,\mev \, ,
&\qquad&
|f_{D_1'}| < 220\,\mev 
\,,
&
\\
&
|g_{D_1}| < 266\,\mev 
\, ,
&&
|g_{D_1'}|< 323\,\mev
\,,
&
\label{eq:uplimnum}
\end{aligned}
\end{equation}
where, to stay on the conservative side, we have increased 
the central values of these bounds by one standard deviation.

By construction, our calculated value  of the $B\to D_1\ell \nu$ total decay width reproduces the measured one in 
\refeq{expBR}. 
On the other hand, our predictions of the $B\to D'_1\ell \nu$ $(\ell =e,\mu)$ width  converted into branching fraction are
\begin{align}
    \text{sol. 1:}\qquad 
    \B(B^+\to \bar{D}^0_1 \ell^+\bar{\nu}_\ell) &= (2.0_{-1.4}^{+2.8})\cdot 10^{-2}
    \,,\\*
    \text{sol. 2:}\qquad 
    \B(B^+\to \bar{D}^0_1\ell^+\bar{\nu}_\ell) &=
    (1.7_{-1.1}^{+2.1})\cdot 10^{-2}
    \,.
\end{align}
They are still compatible with the measured value
$\B(B^+\to \bar{D}^0_1\ell^+\bar{\nu}_\ell) =
    (0.2\pm 0.05)\cdot 10^{-2}$ ~\cite{pdg,Bernlochner:2016bci} but only  within  large uncertainties.

Currently, a very important observable for a semileptonic exclusive $B$ decay is the corresponding lepton-flavour-universality (LFU) ratio. In our case, these ratios are defined as
\begin{align}
    R(D_1^{(\prime)})= \frac{\Gamma(B\to D_1^{(\prime)} \tau \nu)}{\Gamma(B\to D_1^{(\prime)} \ell \nu)}
\end{align}
with $\ell = e,\,\mu$.
Our predictions are the same for both solutions and read
\begin{align}
    R(D_1)  & = 0.10 \pm 0.02 \,,\\
    R(D_1') & = 0.10 \pm 0.03 \,.
\end{align}
The relatively small uncertainties of these results  indicate a partial cancellation of parametric uncertainties
in the ratios of widths.
These predictions are in agreement with the results of Refs.~\cite{Bernlochner:2016bci,Bernlochner:2021vlv}.

\section{Conclusion}

In this paper, we have performed the first calculation of the $B\to D_1$ and the $B \to D_1'$ form factors using QCD light-cone sum rules (LCSRs) with $B$-meson distribution amplitudes (DAs).

In order to disentangle the $D_1$ and $D_1'$ mesons, which are 
close in mass and are both $1^+$ states, we have set up a novel approach that combines QCD
two-point sum rules and LCSRs.
This approach consists in defining two currents that interpolate both $D_1$ and $D_1'$ states
and finding a linear combinations of these currents, which interpolate each of these states individually.
This implies that four decay constants are needed as
inputs to evaluate the $B\to D_1^{(\prime)}$ form factors from the LCSRs with the properly defined currents.
Using three independent two-point sum rules, we have related three decay constants  to the 
fourth one, which we have chosen to be 
$f_{D_1}$ and which is then determined a posteriori using the experimental measurement of the $B\to D_1 \ell \nu$ ($\ell=e,\mu$) branching ratio.

Our new results include analytical expressions for the 
diagonal and non-diagonal two-point correlators of the two interpolating 
currents obtained with local OPE 
and used to derive the two-point sum rules.  In addition, we have calculated the light-cone OPE 
in terms of $B$-meson DAs for the two vacuum-to-$B$ correlators formed by the same interpolating currents with the weak $b\to c$ current and used them to derive LCSRs.

A drawback of our approach is that it yields to a twofold ambiguity, which could be resolved in future with more precise experimental data and/or further independent inputs from theory.
For example, if a lattice QCD result for one of the decay constants would be available, we would neither have a twofold ambiguity nor need to use experimental information.  Hence, it is an important, albeit challenging  task for lattice QCD to isolate the 
narrow $D_1$ resonance and calculate its decay constant of at least one of the interpolating currents.
\\

Our main numerical results include the full set of the $B\to D_1^{(\prime)}$ form factors in the whole semileptonic region. 
Even though these results suffer from large uncertainties caused also by the twofold ambiguity,
we have found that important observables --- such as the lepton flavour universality ratios $R(D_1)$ and $R(D_1')$ ---
can be predicted with a moderate error. It will be interesting to use our results for comparison 
with the  future measurements of semileptonic $B\to D^*\pi\ell \nu_\ell$ decays which are 
accessible, e.g., with the Belle II detector. 
An angular analysis disentangling $S$ and $D$ waves of the $D^*\pi$ final state in these decays 
will allow  to obtain more accurate information on the positions and widths 
of both $1^+$ states. Especially important is to confirm the properties 
of the broad resonance $D_1'$ which are currently less certain. Another type of processes
where our predictions could be used are various nonleptonic $B$ decays with $D_1$ and $D_1^{\prime}$ in the final state
observed as   $D^*\pi$ resonances. Here both LHCb and Belle-II measurements are possible. 
To give just one example: the $B_{(s)}\to D^*\pi\pi(K) $ modes where the predicted decay constants
$f_{D_1^{(\prime)}}$ of the axial current can be used in the  
factorizable approximation to the decay amplitudes.
\\

Future improvements of the approach suggested in this paper are possible and realistic.
They include a more precise calculation of the OPE for both two-point sum rules and LCSRs,
for instance by taking into account perturbative radiative corrections 
and, for the LCSRs, also the subdominant three-particle contributions. Moreover, an 
improvement of the input parameters of $B$-meson DAs, especially of its first inverse moment, 
will also help to pin down the overall parametric uncertainty of our results. 
Achieving this precision of OPE in future  would also demand a  more refined analysis of the
duality thresholds.  One should  use a threshold fixing procedure with the inverse Borel-mass 
differentiation  for each sum rule separately.

Furthermore, a relation to HQET in the context of QCD sum rules remains a very important issue. 
Starting from the correlators with a finite $c$-quark mass considered in this paper 
and expanding them in the powers of inverse mass, could provide
a link to  the heavy-quark limit 
of the resulting sum rules and to get insight into the mixing pattern 
of $j=1/2$ and $j=3/2$ states.
Studying this mixing pattern could also help to resolve the twofold ambiguity of our results.
We plan to return to this problem in future. 

In conclusion, let us mention that the LCSR method with $B$ meson DAs and finite 
$c$ quark mass is universal enough
to  be used also for  the $B$-meson transitions to the other excited
charmed mesons listed in Table~\ref{tab:spectr}, which will be another natural continuation 
of this work.

\subsection*{Acknowledgments}

This research was supported by the Deutsche Forschungsgemeinschaft (DFG, German Research Foundation) under grant 396021762 - TRR 257 “Particle Physics Phenomenology after the Higgs Discovery''. The work of R.M. is supported by the University of Siegen under the Young Investigator Research Group (Nachwuchsforscherinnengruppe) grant.
N.G., A.K., and T.M. acknowledge the hospitality of the Munich Institute for Astro- and Particle Physics (MIAPP) which is funded by the Deutsche Forschungsgemeinschaft (DFG, German Research Foundation) under Germany's Excellence Strategy – EXC-2094 – 390783311. A significant part of the research has been performed at the MIAPP.


\begin{appendix}

\section{$B$-meson distribution amplitudes}
\label{app:BDA}

Following the most complete analysis available for the $B$-meson DAs in Ref.~\cite{Braun:2017liq}\,,
we define the two-particle $B$-meson DAs  as
\begin{multline}
\bra{0} \bar{q}^{\alpha}(x) h_{v}^{\beta}(0) \ket{\bar{B}(v)} =
    -\frac{i f_B m_B}{4} \int^\infty_0 d\omega  \bigg\{
        (1 + \slashed{v}) \bigg[
            \phi_+(\omega) -g_+(\omega) \partial_\lambda \partial^\lambda
            \\*
            +\frac12 \left(\overline{\Phi}_{\pm}(\omega)
            -\overline{G}_\pm(\omega) \partial_\lambda \partial^\lambda\right) 
        \gamma^\rho \partial_\rho
        \bigg] \gamma_5
    \bigg\}^{\beta\alpha} e^{-i l \cdot x}
    \Bigg|_{l=\omega v}
    \,.
    \label{eq:BLCDAs2pt}
\end{multline}
The derivatives $\partial_\mu \equiv \partial/\partial l^\mu$ are understood to act on the hard-scattering kernel.
We have also introduced the notation
\bea
\label{eq:rearr}
\overline{\Phi}_{\pm}(\omega)\equiv \int\limits_0^\omega d\tau
\big(\phi_+(\tau)-\phi_-(\tau)\big)\,,~~ 
\overline{G}_\pm(\omega)\equiv \int\limits_0^\omega d\tau
\big(g_+(\tau)-g_-(\tau)\big)
\,.
\eea
The DAs $\phi_+$, $\phi_-$, $g_+$, and $g_-$ 
are of twist two, three, four, and five, respectively.
For the twist two, three, and four DAs we use the 
exponential model given in Section 5.1 of Ref.~\cite{Braun:2017liq}.
For the twist five DA, which is not specified in Ref.~\cite{Braun:2017liq}, we use its Wandzura-Wilczek approximation given in Eqs.~(A.7)-(A.8) of Ref.~\cite{Gubernari:2018wyi}.

\section{ OPE coefficients and transformation to sum rule}
\label{app:OPEcoeff}

In this appendix we list the coefficients $\C^{(i,k)}_{\L,\psi}$ defined in \refeq{Ink}, which enter the master formula (\ref{eq:MasterFor1}) to compute the OPE of the correlators ${\cal F}^{(i)}_{\L}$:
\begin{align}
\C_{\epsilon_{\mu\nu pq},\phi_+}^{(1,1)}&=-\frac{m_c}{\bar{\sigma}}\,,\\ 
\C_{\epsilon_{\mu\nu pq},\overline{\Phi}_\pm}^{(1,2)}&=\frac{m_c^2}{\bar{\sigma}^2}\,,\\ 
\C_{\epsilon_{\mu\nu pq},g_+}^{(1,2)}&=-\frac{4 m_c}{\bar{\sigma}^2}\,,\\ 
\C_{\epsilon_{\mu\nu pq},g_+}^{(1,3)}&=\frac{8 m_c^3}{\bar{\sigma}^3}\,,\\ 
\C_{\epsilon_{\mu\nu pq},\overline{G}_\pm}^{(1,4)}&=-\frac{24 m_c^4}{\bar{\sigma}^4}\,,\\ 
\C_{g_{\mu\nu},\phi_+}^{(1,1)}&=\frac{i m_c \left(\bar{\sigma}^2 m_B^2-2 \bar{\sigma} m_c m_B+m_c^2-q^2\right)}{2 \bar{\sigma}^2}\,,\\ 
\C_{g_{\mu\nu},\overline{\Phi}_\pm}^{(1,1)}&=-\frac{i m_c^2}{2 \bar{\sigma}^2}\,,\\ 
\C_{g_{\mu\nu},\overline{\Phi}_\pm}^{(1,2)}&=-\frac{i m_c^2 \left(\bar{\sigma}^2 m_B^2-2 \bar{\sigma} m_c m_B+m_c^2-q^2\right)}{2 \bar{\sigma}^3}\,,\\ 
\C_{g_{\mu\nu},g_+}^{(1,1)}&=\frac{2 i m_c}{\bar{\sigma}^2}\,,\\ 
\C_{g_{\mu\nu},g_+}^{(1,2)}&=-\frac{2 i m_c \left(-\bar{\sigma}^2 m_B^2+m_c^2+q^2\right)}{\bar{\sigma}^3}\,,\\ 
\C_{g_{\mu\nu},g_+}^{(1,3)}&=-\frac{4 i m_c^3 \left(\bar{\sigma}^2 m_B^2-2 \bar{\sigma} m_c m_B+m_c^2-q^2\right)}{\bar{\sigma}^4}\,,\\
\C_{g_{\mu\nu},\overline{G}_\pm}^{(1,2)}&=\frac{4 i m_B m_c}{\bar{\sigma}^2}\,,\\ 
\C_{g_{\mu\nu},\overline{G}_\pm}^{(1,3)}&=-\frac{4 i (2 \bar{\sigma} m_B-3 m_c) m_c^3}{\bar{\sigma}^4}\,,\\ 
\C_{g_{\mu\nu},\overline{G}_\pm}^{(1,4)}&=\frac{12 i m_c^4 \left(\bar{\sigma}^2 m_B^2-2 \bar{\sigma} m_c m_B+m_c^2-q^2\right)}{\bar{\sigma}^5}\,,\\ 
\C_{q_\mu q_\nu,\phi_+}^{(1,1)}&=-\frac{2 i (\bar{\sigma}-1) m_c}{\bar{\sigma}}\,,\\ 
\C_{q_\mu q_\nu,\overline{\Phi}_\pm}^{(1,2)}&=-\frac{2 i (\bar{\sigma}-1)^2 m_B m_c}{\bar{\sigma}^2}\,,\\
\C_{q_\mu q_\nu,g_+}^{(1,2)}&=-\frac{8 i (\bar{\sigma}-1) m_c}{\bar{\sigma}^2}\,,\\ 
\C_{q_\mu q_\nu,g_+}^{(1,3)}&=\frac{16 i (\bar{\sigma}-1) m_c^3}{\bar{\sigma}^3}\,,\\ 
\C_{q_\mu q_\nu,\overline{G}_\pm}^{(1,3)}&=-\frac{16 i (\bar{\sigma}-1)^2 m_B m_c}{\bar{\sigma}^3}\,,\\
\C_{q_\mu q_\nu,\overline{G}_\pm}^{(1,4)}&=\frac{48 i (\bar{\sigma}-1)^2 m_B m_c^3}{\bar{\sigma}^4}\,,\\ 
\C_{p_\mu q_\nu,\phi_+}^{(1,1)}&=\frac{i (1-2 \bar{\sigma}) m_c}{\bar{\sigma}}\,,\\ 
\C_{p_\mu q_\nu,\overline{\Phi}_\pm}^{(1,2)}&=-\frac{i \left(2 m_B \bar{\sigma}^2-2 m_B \bar{\sigma}-m_c\right) m_c}{\bar{\sigma}^2}\,,\\ 
\C_{p_\mu q_\nu,g_+}^{(1,2)}&=-\frac{4 i (2 \bar{\sigma}-1) m_c}{\bar{\sigma}^2}\,,\\ 
\C_{p_\mu q_\nu,g_+}^{(1,3)}&=\frac{8 i (2 \bar{\sigma}-1) m_c^3}{\bar{\sigma}^3}\,,\\ 
\C_{p_\mu q_\nu,\overline{G}_\pm}^{(1,3)}&=-\frac{16 i (\bar{\sigma}-1) m_B m_c}{\bar{\sigma}^2}\,,\\ 
\C_{p_\mu q_\nu,\overline{G}_\pm}^{(1,4)}&=\frac{24 i \left(2 m_B \bar{\sigma}^2-2 m_B \bar{\sigma}-m_c\right) m_c^3}{\bar{\sigma}^4}\,,\\ 
\C_{\epsilon_{\mu\nu pq},\overline{\Phi}_\pm}^{(2,1)}&=\frac{1}{\bar{\sigma}}\,,\\ 
\C_{\epsilon_{\mu\nu pq},\overline{G}_\pm}^{(2,2)}&=\frac{4}{\bar{\sigma}^2}\,,\\ 
\C_{\epsilon_{\mu\nu pq},\overline{G}_\pm}^{(2,3)}&=-\frac{8 m_c^2}{\bar{\sigma}^3}\,,\\ 
\C_{g_{\mu\nu},\overline{\Phi}_\pm}^{(2,1)}&=-\frac{i \left(\bar{\sigma}^2 m_B^2-2 \bar{\sigma} m_c m_B+m_c^2-q^2\right)}{2 \bar{\sigma}^2}\,,\\ 
\C_{g_{\mu\nu},g_+}^{(2,1)}&=\frac{4 i m_B}{\bar{\sigma}}\,,\\ 
\C_{g_{\mu\nu},\overline{G}_\pm}^{(2,1)}&=-\frac{2 i}{\bar{\sigma}^2}\,,\\ 
\C_{g_{\mu\nu},\overline{G}_\pm}^{(2,2)}&=-\frac{2 i \left(\bar{\sigma}^2 m_B^2-2 \bar{\sigma} m_c m_B-m_c^2-q^2\right)}{\bar{\sigma}^3}\,,\\
\C_{g_{\mu\nu},\overline{G}_\pm}^{(2,3)}&=\frac{4 i m_c^2 \left(\bar{\sigma}^2 m_B^2-2 \bar{\sigma} m_c m_B+m_c^2-q^2\right)}{\bar{\sigma}^4}\,,\\ 
\C_{q_\mu q_\nu,\phi_+}^{(2,1)}&=-\frac{2 i (\bar{\sigma}-1) ((\bar{\sigma}-1) m_B+m_c)}{\bar{\sigma}}\,,\\ 
\C_{q_\mu q_\nu,\overline{\Phi}_\pm}^{(2,1)}&=\frac{4 i (\bar{\sigma}-1)}{\bar{\sigma}}\,,\\ 
\C_{q_\mu q_\nu,\overline{\Phi}_\pm}^{(2,2)}&=-\frac{2 i (\bar{\sigma}-1) m_c ((\bar{\sigma}-1) m_B+m_c)}{\bar{\sigma}^2}\,,\\ 
\C_{q_\mu q_\nu,g_+}^{(2,2)}&=-\frac{8 i (\bar{\sigma}-1) (2 (\bar{\sigma}-1) m_B+m_c)}{\bar{\sigma}^2}\,,\\ 
\C_{q_\mu q_\nu,g_+}^{(2,3)}&=\frac{16 i (\bar{\sigma}-1) m_c^2 ((\bar{\sigma}-1) m_B+m_c)}{\bar{\sigma}^3}\,,\\ 
\C_{q_\mu q_\nu,\overline{G}_\pm}^{(2,2)}&=\frac{16 i (\bar{\sigma}-1)}{\bar{\sigma}^2}\,,\\ 
\C_{q_\mu q_\nu,\overline{G}_\pm}^{(2,3)}&=-\frac{16 i (\bar{\sigma}-1) m_c ((\bar{\sigma}-1) m_B+2 m_c)}{\bar{\sigma}^3}\,,\\ 
\C_{q_\mu q_\nu,\overline{G}_\pm}^{(2,4)}&=\frac{48 i (\bar{\sigma}-1) m_c^3 ((\bar{\sigma}-1) m_B+m_c)}{\bar{\sigma}^4}\,,\\ 
\C_{p_\mu q_\nu,\phi_+}^{(2,1)}&=-\frac{2 i (\bar{\sigma}-1) (\bar{\sigma} m_B+m_c)}{\bar{\sigma}}\,,\\ 
\C_{p_\mu q_\nu,\overline{\Phi}_\pm}^{(2,1)}&=4 i-\frac{3 i}{\bar{\sigma}}\,,\\ 
\C_{p_\mu q_\nu,\overline{\Phi}_\pm}^{(2,2)}&=-\frac{2 i (\bar{\sigma}-1) m_c (\bar{\sigma} m_B+m_c)}{\bar{\sigma}^2}\,,\\ 
\C_{p_\mu q_\nu,g_+}^{(2,2)}&=-\frac{8 i (\bar{\sigma}-1) (2 \bar{\sigma} m_B+m_c)}{\bar{\sigma}^2}\,,\\ 
\C_{p_\mu q_\nu,g_+}^{(2,3)}&=\frac{16 i (\bar{\sigma}-1) m_c^2 (\bar{\sigma} m_B+m_c)}{\bar{\sigma}^3}\,,\\ 
\C_{p_\mu q_\nu,\overline{G}_\pm}^{(2,2)}&=\frac{4 i (4 \bar{\sigma}-3)}{\bar{\sigma}^2}\,,\\ 
\C_{p_\mu q_\nu,\overline{G}_\pm}^{(2,3)}&=-\frac{8 i m_c \left(2 m_B \bar{\sigma}^2-2 m_B \bar{\sigma}+4 m_c \bar{\sigma}-3 m_c\right)}{\bar{\sigma}^3}\,,\\
\C_{p_\mu q_\nu,\overline{G}_\pm}^{(2,4)}&=\frac{48 i (\bar{\sigma}-1) m_c^3 (\bar{\sigma} m_B+m_c)}{\bar{\sigma}^4}\,.
\end{align}

\section{$B \to D_1^{(\prime)} \ell \bar\nu$ total decay width}
\label{app:width}

The differential distribution for the semileptonic decay $B \to D_1^{(\prime)} \ell \bar\nu$ with respect to the lepton-neutrino invariant mass square $q^2$ can be written as
\begin{equation}
\begin{aligned}
\frac{d\Gamma (B\to D_1^{(\prime)}\ell\nu_\ell)}{dq^2} 
&= 
\frac{G_F^2 |V_{cb}|^2}{192 m_B^3 \pi^3}
q^2  \lambda^{1/2}(m_B^2,m_{D_1^{(\prime)}}^2,q^2) \left(1- \frac{m_\ell^2}{q^2} \right)^2 
\\ & \times
\bigg[ \left(1+ \frac{m_\ell^2}{2q^2} \right)^2 \{ |H_+^{(\prime)}|^2 + |H_-^{(\prime)}|^2 + |H_0^{(\prime)}|^2\}  + \frac{3 m_\ell^2}{2 q^2}  |H_t^{(\prime)}|^2 \bigg] \,.
\label{eq:D1width}
\end{aligned}
\end{equation}
Here, the helicity amplitudes are
\begin{align}
    H_{+,-}^{(\prime)}  &= i(m_B + m_{D_1^{(\prime)}}) V_1^{BD_1^{(\prime)}}  (q^2) \mp \frac{i\lambda^{1/2}(m_B^2,m_{D_1^{(\prime)}}^2,q^2)}{m_B + m_{D_1^{(\prime)}}} A^{BD_1^{(\prime)}} (q^2)\,, \\ 
H_0^{(\prime)} &= i \frac{m_B+m_{D_1^{(\prime)}}}{2 m_{D_1^{(\prime)}} \sqrt{q^2}} \left( ( m_{D_1^{(\prime)}}^2 +q^2 - m_B^2) V_1^{BD_1^{(\prime)}}(q^2) + \frac{\lambda(m_B^2,m_{D_1^{(\prime)}}^2,q^2)}{(m_B + m_{D_1^{(\prime)}})^2} V_2^{BD_1^{(\prime)}}(q^2) \right)\,, \\
H_t^{(\prime)}&= -i\frac{\lambda^{1/2}(m_B^2,m_{D_1^{(\prime)}}^2,q^2)}{\sqrt{q^2}} V_0^{BD_1^{(\prime)}}\,.
\end{align}
Equation (\ref{eq:D1width}) coincides with the formulas given in~\cite{Bernlochner:2016bci}.
The relations between their HQET basis of form factors and our basis are
\begin{align}
A^{BD_1}=&\frac{i\left(m_B + m_{D_1} \right)}{2\sqrt{m_B m_{D_1}}} f_A\,, \\
V_1^{BD_1}=& \frac{i\sqrt{m_B m_{D_1}}}{m_B + m_{D_1 }} f_{V_1}\,, \\
V_2^{BD_1}=& -\frac{i\left(m_B + m_{D_1} \right)}{2\sqrt{m_B m_{D_1}}} \left[ f_{V_3} + \frac{m_{D_1}}{m_B} f_{V_2} \right]\,, \\
V_0^{BD_1}=& \frac{i}{2\sqrt{m_B m_{D_1}}}\left[ m_B f_{V_1} + \frac{m_B^2-m_{D_1}^2+q^2}{2m_B} f_{V_2} + \frac{m_B^2-m_{D_1}^2-q^2}{2m_{D_1}} f_{V_3} \right]\,, \\
A^{BD_1^\prime}=&\frac{i\left(m_B + m_{D_1^\prime} \right)}{2\sqrt{m_B m_{D_1^\prime}}} g_A\,, \\
V_1^{BD_1^\prime}=& \frac{i\sqrt{m_B m_{D_1^\prime}}}{m_B + m_{D_1^\prime }} g_{V_1}\,, \\
V_2^{BD_1^\prime}=& -\frac{i\left(m_B + m_{D_1^\prime} \right)}{2\sqrt{m_B m_{D_1^\prime}}} \left[ g_{V_3} + \frac{m_{D_1^\prime}}{m_B} g_{V_2} \right]\,, \\
V_0^{BD_1^\prime}=& \frac{i}{2\sqrt{m_B m_{D_1^\prime}}}\left[ m_B g_{V_1} + \frac{m_B^2-m_{D_1^\prime}^2+q^2}{2m_B} g_{V_2} + \frac{m_B^2-m_{D_1^\prime}^2-q^2}{2m_{D_1^\prime}} g_{V_3} \right]\,. 
\end{align}

\end{appendix}

\end{document}

%% file: macros.tex
\usepackage{multirow}
\usepackage{booktabs}

\usepackage{hyperref}
\usepackage{placeins}
\usepackage{verbatim}

\usepackage[normalem]{ulem}
\usepackage{cancel}

\usepackage{float} 

\def\bc{\begin{center}}
\def\ec{\end{center}}
\def\be{\begin{equation}}
\def\ee{\end{equation}}
\def\bea{\begin{eqnarray}}
\def\eea{\end{eqnarray}}

\def\gev{\ensuremath{\mathrm{Ge\kern -0.1em V}}}
\def\mev{\ensuremath{\mathrm{Me\kern -0.1em V}}}

\def \Im{\text{Im}}

\newcommand{\OPE}{\text{OPE}}
\newcommand{\HAD}{\text{had}}

\newcommand{\pert}{\text{pert}}
\newcommand{\cond}{\text{cond}}
\newcommand{\cont}{\text{cont}}
\renewcommand{\th}{\text{th}}

\newcommand{\MeV}{\text{\,MeV}}
\newcommand{\GeV}{\text{\,GeV}}
\newcommand{\B}{\mathcal{B}}
\newcommand{\C}{\mathcal{C}}
\newcommand{\E}{\mathcal{E}}

\newcommand{\T}{\mathcal{T}}
\renewcommand{\L}{\mathcal{L}}

\definecolor{darkgreen}{RGB}{30,150,30}

\newcounter{TODO}

\makeatletter

\def\nig{\@ifstar\@@nig\@nig}
\newcommand{\@nig}[1]{\textcolor{brown}{[\textbf{NG:} #1]}}
\newcommand{\@@nig}[1]{\textcolor{brown}{#1}}

\makeatother

\newcommand{\refapp}[1]{Appendix~\ref{app:#1}}

\newcommand{\refeq}[1]{Eq.~(\ref{eq:#1})}
\newcommand{\refeqs}[2]{Eqs.~(\ref{eq:#1})-(\ref{eq:#2})}
\newcommand{\refsec}[1]{Section~\ref{sec:#1}}
\newcommand{\reftab}[1]{Table~\ref{tab:#1}}